\documentclass[a4paper]{article}

\usepackage{amsmath}
\usepackage{amssymb}
\usepackage{color}
\usepackage{graphicx}
\usepackage{authblk}

\begin{document}

\title{The spectra and composition of Ultra High Energy Cosmic Rays 
 and the measurement of the proton-air cross section}

\author[1]{\normalsize Paolo Lipari}

\affil[1]{\footnotesize INFN, Sezione Roma ``Sapienza''}

\maketitle

\begin{abstract}
 The shape of the longitudinal development of the showers generated
 in the atmosphere by very high energy cosmic ray particles 
 encodes information about the mass composition of the flux,
 and about the properties of hadronic interactions that control the shower development.
 Studies of the energy dependence of the average and width of the depth of maximum
 distribution of showers
 with $E \gtrsim 10^{17.3}$~eV measured by the Pierre Auger Observatory,
 suggest, on the basis of a comparison with current models, that the composition of the
 cosmic ray flux undergoes a very important evolution, first becoming lighter and then rapidly
 heavier. These conclusions, if confirmed, would have profound and very
 surprising implications for our understanding of the high energy astrophysical sources.
 Studies of the shape of the depth of maximum distribution in the same energy range
 have been used by Auger and by the Telescope Array Collaboration
 to measure the interaction length of protons in air, a quantity that allows
 to estimate the $pp$ cross sections for values of $\sqrt{s}$
 well above the LHC range.
 In this paper we argue that it is desirable
 to combine the studies of the cosmic ray composition
 with those aimed at the measurement of the $p$--air cross section.
 The latter allow to obtain estimates for the fraction of protons in the flux
 that can be of great help in decoding the composition and its energy dependence.
 Studies that consider multiple parameters to characterize the depth of maximum distributions
 also offer the possibility to perform more sensitive tests of the validity of the models
 used to describe high energy showers.
\end{abstract}


\section{Introduction}
\label{sec:intro}
The study of ultra high energy cosmic rays (UHECR) with $E \gtrsim 10^{17}$~eV
is essential to develop an understanding of high energy sources in the universe.
At present the shape of the all--particle spectrum in this energy range is reasonably well
measured, thanks to the fact that
fluorescence light observation \cite{LetessierSelvon:2011dy}
of cosmic ray (CR) showers allow a calorimetric, in good approximation model independent
measurement of the primary particle energy. 
The composition of the cosmic ray flux remains however more poorly determined.
Information about the mass of the primary particle is encoded in the shape
of the longitudinal profile of the showers, however the
determination of the composition from the data
is difficult because the development of the CR showers
also depend on the properties of hadronic interactions,
that are not well understood. Uncertainties associated to the description
of hadronic interactions are the main limitation for the program
to determine the CR composition from fluorescence light observations.

It is generally recognized that QCD gives the fundamental Lagrangian that describes
hadronic interactions in terms of quark and gluon fields,
however at present we are not able to use the theory to compute
all phenomenologically relevant quantities from first principles. 
Experimental studies at accelerators have provided a large amount of information
that allows to model with reasonable accuracy interactions in a broad range
of energies, but the study of UHECR requires an extrapolation (up to $\sqrt{s} \simeq 430$~eV)
from the highest energy results (obtained at LHC at $\sqrt{s} = 13$~TeV).
This of course also offers the possibility to use CR observations
to perform experimental studies of hadronic interactions above the LHC range.

The problem here is that these two goals
appear to be in conflict with each other.
On one hand the measurement of the CR composition
requires a comparison of the data with models 
that must include a description of the properties of hadronic interactions,
and on the other hand it is problematic to extract information
about hadronic interactions from CR data,
because the mass composition of the ``beam'' is not known.
An attractive possibility is to use self--consistency
in the simultaneous study of several different observables
to extract information on both the CR composition and hadronic interactions.

In recent years very large aperture cosmic ray detectors such as
the Pierre Auger Observatory in Argentina \cite{Abraham:2008ru}
and Telescope Array in the USA \cite{AbuZayyad:2012ru}
have collected large statistics of events in the UHE range.
Interpretations of the (higher statistics) Auger observations
based on a comparison of the data with 
Monte Carlo simulations that include detailed descriptions of hadronic interactions
suggest that the CR composition is continuosly changing with energy
\cite{Aab:2014aea,Aab:2016zth,Castellina:2019huz,Aab:2020rhr},
first becoming lighter (for $E \lesssim 10^{18.3}$~eV) and then
very rapidly heavier, with indications that the spectra of different
elements have sharp cutoffs at maximum energies that are roughly proportional to the
nucleus electric charge $Z$.
These results, if confirmed would have profound and very
surprising implications for high energy
astrophysics, and it is very important to confirm (or falsify)
them with additional studies.
Studies on the cosmic ray composition have also be performed
by the Telescope Array detector
\cite{Hanlon:2019onl,Bergman:2020zac}. The consistency
of the Auger and Telescope Array results is under careful study
(see the discussions in \cite{Unger:2015rzh,Hanlon:2018dhz}).

The interpretation of the Auger data in terms of an energy dependent
composition is based on a comparison of the data with Monte Carlo
models, and the detailed form of the energy dependence of the composition
is model dependent.
It is therefore very important to try to validate the models used
in these studies, and if possible narrow the range of theoretical uncertainties.

The observations of the longitudinal profiles of UHE showers
have also been used to obtain measurements
of the $p$--air inelastic (production) cross section
\cite{Baltrusaitis:1984ka,auger-sigma-2012,Abbasi:2015fdr,Abbasi:2020chd},
in an energy range ($\sqrt{s} \simeq 30$--95~TeV for
nucleon--nucleon collisions) that is above the maximum energy
obtained at LHC. From these measurements it is possible to
infer the cross sections for $pp$ collisions using well established
theoretical concepts that relate the properties
of hadron--nucleon and hadron--nucleus collisions \cite{Glauber:1970jm}.

These cross section measurements adopt
a method pioneered by the Fly's Eye collaboration \cite{Baltrusaitis:1984ka}
that relates the slope of the exponential tail of the distribution
of depth of maximum of the showers to the interaction length of protons in air.
The authors of these works argue that the results are
(in good approximation) model independent,
that is insensitive to other
properties of hadronic interactions such as the average multiplicity or
the inclusive spectra of final state particles,
and also insensitive to the exact composition of the CR flux,
as long as protons are a significant component.

The main goal of this paper is to argue that there are
significant advantages if these two type of studies,
that use the same data to achieve different goals
(the measurement of the CR mass composition and the
$p$--air cross section) are combined and performed together.
The measurement of the proton cross section, must after all
necessarily identify a proton component, and it is natural to
include this information in the study of the CR composition.

A combined study of composition and of the
shape of proton--induced showers offers the possibility
to reduce the systematic errors for both measurements.
In addition, and perhaps even more important,
a study where more observables are considered simultaneously allows
more stringent tests for the validity of the Monte Carlo codes.
It is methodologically important not to discard a priori the logical
possibility that our current understanding of hadronic interactions
is incomplete and that new phenomena, not detectable at lower energy
with accelerator experiments, are present in the UHECR range and distort
the interpretation of the data.
These phenomena can be revealed in multi--parameter studies of
the shower properties.

This work is organized as follows,
in the next section we review some relevant UHECR observations.
Section~\ref{sec:interpretation} discusses the evolution of the cosmic ray composition
that can be inferred comparing Monte Carlo models with the Auger measurements
of the average and width of the depth of maximum of the showers.
The following section discusses the measurements of the proton--air cross section
obtained from the study of the shape of the $X_{\rm max}$ distributions, and
discusses how this information also allows to estimate the fraction of protons
in the cosmic ray flux. The last section contains some final considerations.

\section{Observations of Ultra High Energy Cosmic Rays}
\label{sec:observations}

\subsection{All--particle Energy spectrum}
Cosmic Rays at very high energy can be observed using two
different techniques. 
 In the ``surface array technique''
a network of sensors at the surface of the Earth observes a fraction
of the particles in the shower that reach the ground.
In the ``fluorescence technique'' the photons isotropically emitted by
nitrogen molecules excited by the passage of a CR shower are observed
by telescopes at the ground to reconstruct the longitudinal
profile $N(X)$, that is the number of charged particles at column density $X$.

Integrating the longitudinal profile $N(X)$ over all $X$
and multiplying by the average energy loss $\langle dE/dX\rangle$
of relativistic charged particles in air, one obtains
the energy of the shower dissipated as ionization in the atmosphere,
a quantity that accounts for most of the primary particle initial energy.
Including corrections for the ``invisible energy''
carried by neutrinos and for the energy dissipated in the ground,
the measurement of the longitudinal profile 
yields then an estimate of the energy of the primary CR particle
that is in good approximation independent from from its mass, and from
the modeling of the shower development.
Cosmic ray observatories such as Auger and Telescope Array are
hybrid detectors that use both techniques, and the
fluorescence light observations can then also be used
to calibrate the data of the surface array, allowing a determination of
the all--particle spectrum with higher statistics.

\begin{figure}[bt]
\begin{center}
\includegraphics[width=16.0cm]{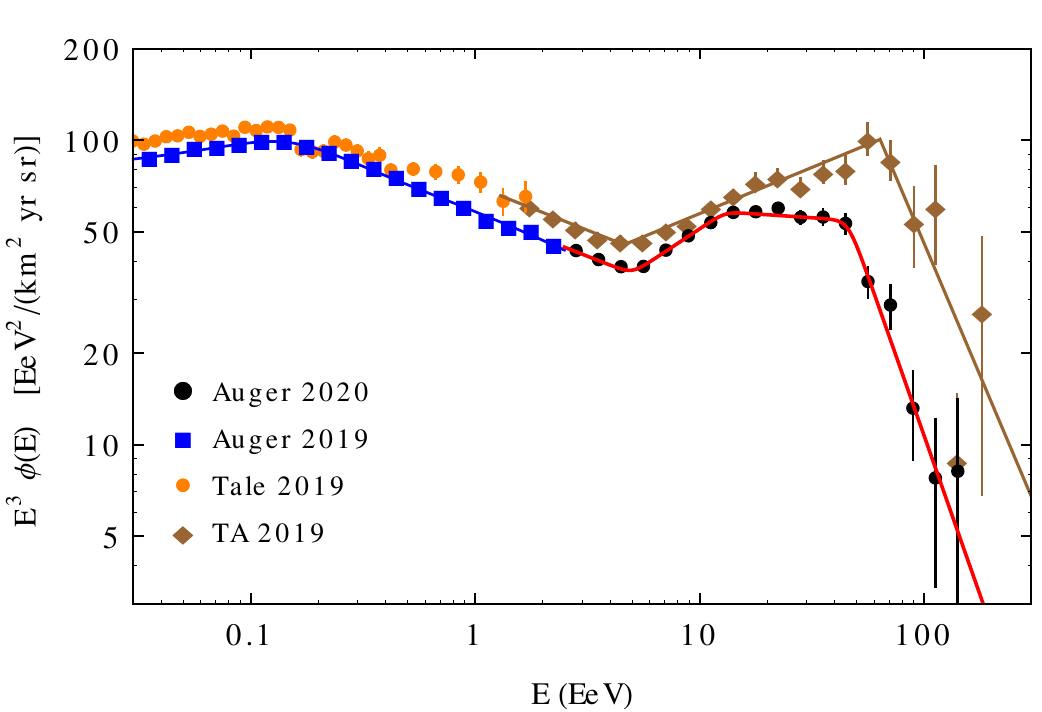}
\end{center}
\caption {\footnotesize
All particle energy spectrum of very high energy cosmic rays.
The measurements are by Auger \cite{Aab:2020gxe,verzi:2019u},
Telescope Array \cite{ivanov_ta_icrc2019}
and TALE \cite{Abbasi:2018xsn}.
The lines are fits to the data reported in the original publications.
\label{fig:spectrum_uhecr} }
\end{figure}

Fig.~\ref{fig:spectrum_uhecr} show measurements of the all--particle spectrum
obtained by the Pierre Auger Observatory
(taken from \cite{Aab:2020gxe} for $E > 2.5 \times 10^{18}$~eV and from
\cite{verzi:2019u} at lower energy), 
and by Telescope Array \cite{ivanov_ta_icrc2019}
and TALE \cite{Abbasi:2018xsn}. The results are in reasonable
good agreeement, with some discrepancies emerging only at the highest energies.
The main spectral features are:
\begin{itemize}
\item[(A)] A softening around $E \simeq 1.5 \times 10^{17}$~eV,
 commonly called the ``second knee''.
\item[(B)] A marked hardening
 commonly called the ``ankle'' observed by both Auger
 and Telescope Array at 
 $E \simeq 5.0 \times 10^{18}$~eV 
\item [(C)] In the energy decade between the second knee and the
 ``ankle'' the all particle spectrum is well described
 by a simple power law. The spectral index is estimated by Auger in
 the entire energy interval \cite{verzi:2019u}
 as $\gamma_1 = 3.27 \pm 0.05$, and for $E > 2.5 \times 10^{18}$~eV in \cite{Aab:2020gxe}
 as $\gamma_1 \simeq 3.29 \pm 0.02 \pm 0.1$.
 The best fit for Telescope Array \cite{ivanov_ta_icrc2019} is
 $\gamma_1 \simeq 3.28 \pm 0.02$. 

\item [(D)] A strong suppression of the flux is observed at
 $E \approx 5 \times 10^{19}$~eV.
\item [(E)] The Auger collaboration
 \cite{Aab:2020gxe,verzi:2019u} has fitted the spectral shape
 between the ankle and the high energy suppression as
 a broken power law, with a spectral break at
 $E^* \simeq (13\pm 1 \pm 2) \times 10^{18}$~eV, and exponents
 $\gamma_2 \simeq 2.51 \pm 0.03 \pm 0.05$ and
 $\gamma_3 \simeq 3.05 \pm 0.04 \pm 0.10$
 in the lower and higher energy range.
 The spectrum measured by Telescope Array in the same range in consistent
 with an unbroken power law of slope $2.68 \pm 0.02$.
\end{itemize}
To understand the origin of the spectral features in the all--particle spectrum,
it seems vital to determine also the composition as a function of energy.

\subsection{Depth of maximum distributions}
While the integral of a shower longitudinal profile is entirely determined
by the primary particle
energy, its {\em shape} depends on the mass number $A$ of the particle, and on the
properties of hadronic interactions.
A shower profile can be characterized by
several parameters (see for example the discussion in \cite{Lipari:2016dzt})
however essentially all studies
until now have relied entirely of the most characteristic one,
the depth of maximum $X_{\rm max}$ that is the
column density where the profile has its maximum.

Shower development is a stochastic process
where fluctuations are large and important.
Therefore the showers generated by a primary particles
of a fixed energy and mass number have a broad
distribution of depth of maximum $F_A (X_{\rm max}, E)$.
Simulation with Monte Carlo codes allow to construct predictions
for the $X_{\rm max}$ distributions that can then be compared to the data
to infer the CR mass composition.

\begin{figure}[bt]
\begin{center}
\includegraphics[width=12.0cm]{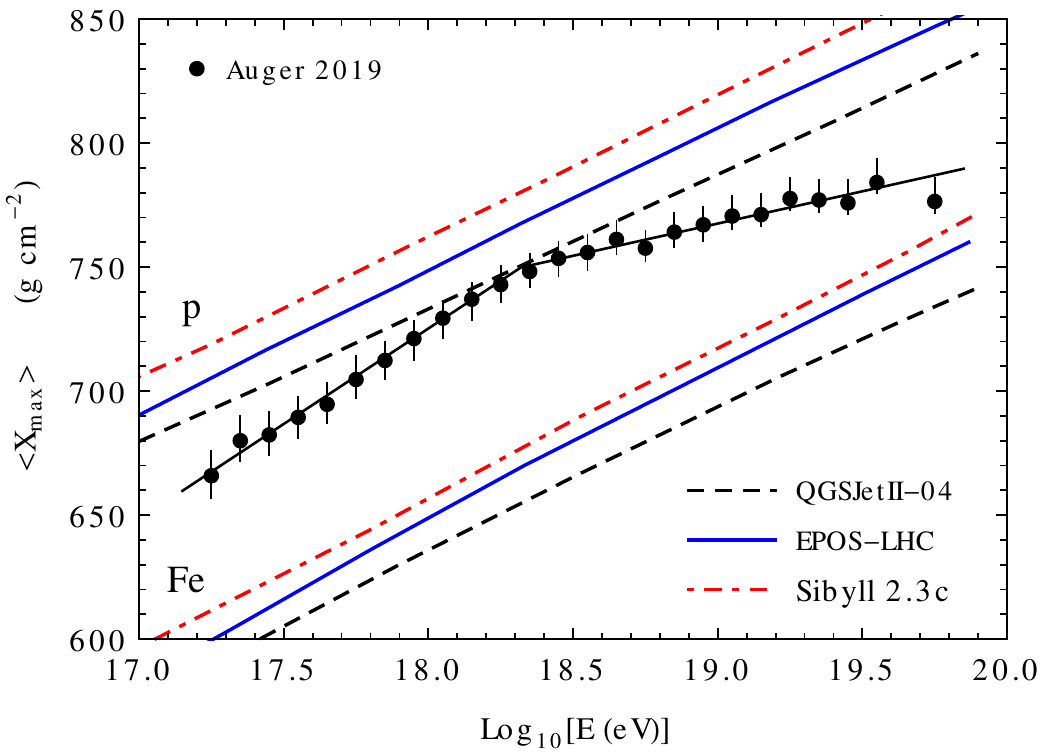}
~~~
\vspace{0.9 cm}

\includegraphics[width=12.0cm]{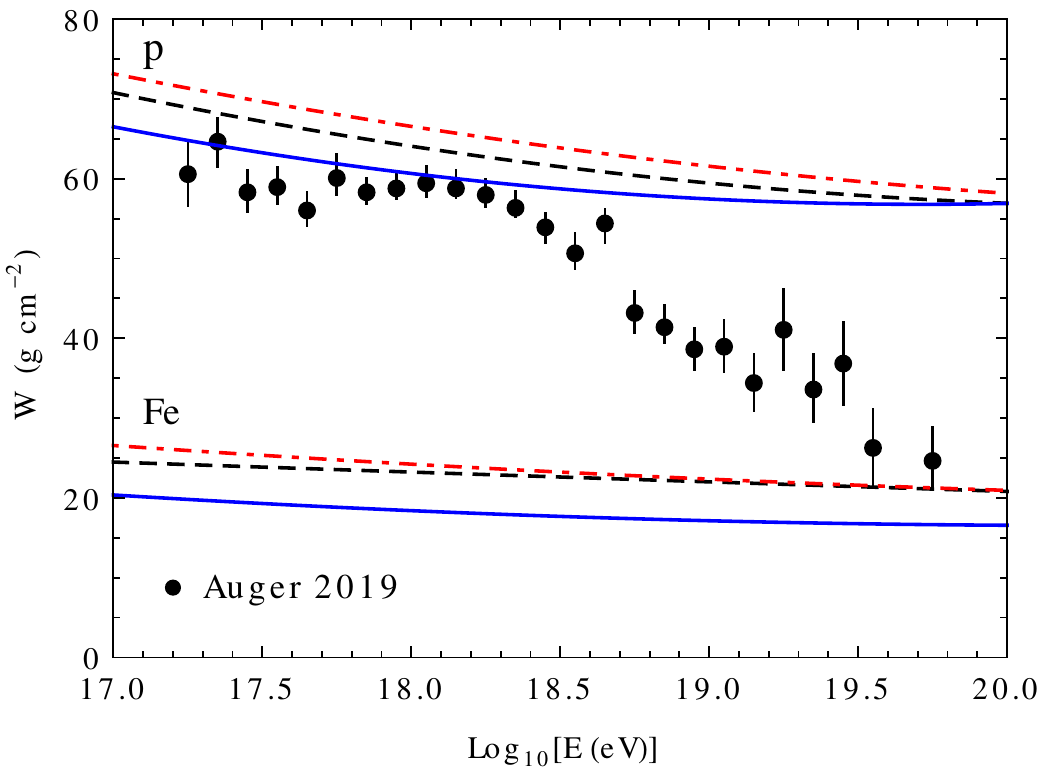}
\end{center}
\caption {\footnotesize Measurements of the average 
 $\langle X_{\rm max} \rangle$ (top--panel)
 and dispersion $W= [\langle X_{\rm max}^2 \rangle - \langle X_{\rm max} \rangle^2]^{1/2}$
 (bottom panel) of the depth of maximum distributions measured by the
 Pierre Auger Observatory 
 in different energy intervals \cite{yushkov-icrc2019}.
 The predictions
 for protons and iron nuclei particles are calculated using
 the hadronic models
 QGSJet~II--04 \cite{Ostapchenko:2013pia},
 EPOS--LHC \cite{Pierog:2013ria} 
 and Sibyll~2.3c \cite{Fedynitch:2018cbl}.
\label{fig:auger_long} }
\end{figure}

Measurements of the average $\langle X_{\rm max} \rangle$ and width
$W = \sqrt{\langle X_{\rm max}^2\rangle - \langle X_{\rm max}\rangle^2}$ of the depth of maximum
for the showers detected by Auger \cite{yushkov-icrc2019} in different energy bins 
are shown in Fig.~\ref{fig:auger_long} together with
prediction for pure compositions of protons and iron nuclei calculated
for three models for shower development:
QGSJet II--04 \cite{Ostapchenko:2013pia}, EPOS--LHC \cite{Pierog:2013ria} 
and Sibyll 2.3c \cite{Fedynitch:2018cbl}.

Measurements of $\langle X_{\rm max} \rangle$ and $W$ have also been
obtained by Telescope Array \cite{Abbasi:2018nun}. The results of the two experiments
however cannot be directly and easily compared to each other
because the measurements made public have not been corrected for
significant detector acceptance effects. 
The question of the consistency between the results on composition of
the two experiments has been the objects of detailed joint studies
\cite{Unger:2015rzh,Hanlon:2018dhz} that will not be reviewed in the present work,
that in the following will concentrate on the interpretation of the
higher statistics Auger data.

Inspecting Fig.~\ref{fig:auger_long} one can see that the three models
have predictions for the average and width of the
depth of maximum distributions that have some important similarities:
\begin{enumerate}
\item The average $X_{\rm max}$ for protons in good approximation
 grows linearly with $\log E$, with an elongation rate
 $D(E) = d\langle X\rangle / d\log E$ that is approximately energy 
 independent. For protons at $E \simeq 10^{18.5}$ the three models
 have elongation rates that are very similar: 54.0, 56.7 and 57.2 g/(cm$^2$~decade)
(for QGSJet II--04, EPOS--LHC and Sibyll 2.3c respectively).

\item The absolute value of the depth of maximum for protons
 is however model dependent, with predictions (always at $E \simeq 10^{18.5}$)
 for the three models: $\langle X_p \rangle = 760$, 778 and 790~g/cm$^2$.

\item The average depth of maximum for iron nuclei 
 has approximately the same energy dependence as for protons, so that 
 the difference in $\langle X_{\rm max} \rangle$ between proton
 and iron showers is approximately constant,
and has only a small model dependence, with value
 $\langle X_p \rangle - \langle X_{\rm Fe} \rangle \simeq 90$--100~g/cm$^2$.
 This can be understood noting that 
 the shower generated by a nucleus of energy $E$ and mass number
 $A$, in good approximation can be described as the superposition of
 $A$ nucleon showers of energy $E/A$.
 The energy and mass dependences of the average depth of maximum
 can then be summarized with a simple equation that is not exact,
 but captures the main properties of the current models:
 \begin{equation}
\langle X_A (E) \rangle \simeq \langle X_p (E_0) \rangle + D_0 \, \log
\left (\frac{E}{A \; E_0} \right ) 
\label{eq:theory_simple}
 \end{equation}
where $E_0$ is an arbitrary reference energy, $\langle X_p (E_0) \rangle$ is the average
depth of maximum for protons at this energy, and $D_0$ is a theoretical ``elongation rate''
that is approximately energy independent. 

\item The width of the $X_{\rm max}$ distributions changes only slowly
 with energy, and decreases with $A$ being of order 60~g/cm$^2$ for
 protons and of order 20~g/cm$^2$ for iron with only a weak
 model dependence.
 First order approximation of the $A$ dependence of the width are:
 $W_A^2 \simeq W_p^2 ~A^{-0.5}$, or $W_A^2 \simeq W_p^2 \; [1 - a \, \log A + b \, (\log A)^2]$
 (with $a$ and $b$ adimensional positive constants).
\end{enumerate}

Using the approximation of Eq.~(\ref{eq:theory_simple})
one finds that for a mixed composition
the average of the distribution is:
\begin{equation}
 \langle X_{\rm max} \rangle \simeq 
 \langle X_p \rangle - D_0 \; \langle \log A \rangle
\label{eq:simple_average}
\end{equation}
and depends linearly on the average of the logarithm of the mass number of the particles
that form the flux, while the width takes the form:
\begin{equation}
W^2 \simeq \left \langle W_A^2 \right \rangle + D_0^2 ~\sigma_{\log A}^2
\label{eq:simple_width}
\end{equation}
where the first term is the average of the widths of the distributions
for each mass component, and in the second term $\sigma_{\log A}$
is the r.m.s. of the $\log A$ distribution.
It is instructive to consider the simple case of a spectrum formed
by two componente of mass $A_1$ and
$A_2$, when Eq.~(\ref{eq:simple_width}) can be rewritten as:
\begin{equation}
 W^2 \simeq f_1\; W_{A_1}^2 + (1-f_1) \; W_{A_2}^2 + D_0^2 \, f_1 \, (1-f_1) [\log A_1 - \log A_2]^2
\label{eq:simple_width1}
\end{equation}
where $f_1$ is the fraction of the flux of the $A_1$ component.
If the two mass numbers $A_1$ and $A_2$ are sufficiently different, the last term
in the equation, that takes into account of the fact that the distributions
of the two components are centered on different average values, becomes dominant.
For example, combining protons with iron,
and using the values for $W_p$ and $W_{\rm Fe}$ of the current Monte Carlo models,
one finds that the dispersion of the mixed composition
is {\em larger} that the width for a pure proton composition
if $f_p \gtrsim 0.31$, with the broadest distribution
($W \approx 1.15~W_p$) obtained for $f_p \approx 0.65$.
Similarly, combining proton with silicon, the width of the mixture is
broader than for pure protons when $f_p \gtrsim 0.43$, 
with the broadest distribution ($W \approx 1.08~W_p$) obtained for $f_p \simeq 0.71$.

In \cite{yushkov-icrc2019} the elongation rate $D_{\rm data}$ 
of the Auger measurement is fitted to the values $77 \pm 2$
and $26 \pm 2$ [in units g/(cm$^{2}$~decade)]
below and above the energy $E^* \simeq 10^{18.3}$~eV.
Since the elongation rate of the models for a constant compositions
$D_{0}$ is predicted to be in the range 54--61~g/(cm$^2$~decade),
one has to conclude that if the models are correct,
the CR composition must change with energy,
becoming gradually lighter in the lower energy range,
and then gradually heavier at higher energy.

The measurements of the width of the $X_{\rm max}$ distribution are
a very important constraint on the evolution of the composition.
Below $E^* \simeq 10^{18.3}$~eV the width is approximately
constant, with a value $W \simeq 60$~g/cm$^2$ that is consistent with 
the prediction for a pure proton composition.
At higher energy the width decreases monotonically, reaching
a value of order 30~g/cm$^2$ at $E \approx 10^{19.5}$~eV.

The CR composition at a give energy is determined by the set $\{f_A\}$
that give the fractions of the flux in nuclei of mass number $A$.
Given these mass fractions, and a model for shower development that predicts
the values of $\langle X_A \rangle$ and $W_A$,
it is straightforward to compute the expected average and width
of the depth of maximum distribution. 
The inverse mapping however has not in general a unique solution, 
because different compositions can result in $X_{\rm max}$ distributions that
have identical average and dispersion.
It has been show \cite{Kampert:2012mx} that in a reasonably good approximation
there is a one to one mapping between 
$\{\langle X_{\rm max} \rangle, W\}$ and the
the pair of parameters $\{\langle \log A \rangle , \sigma^2_{\log A} \}$,
that give the average and r.m.s. of the $\log A$ distribution.

\begin{figure}[bt]
\begin{center}
\includegraphics[width=9.0cm]{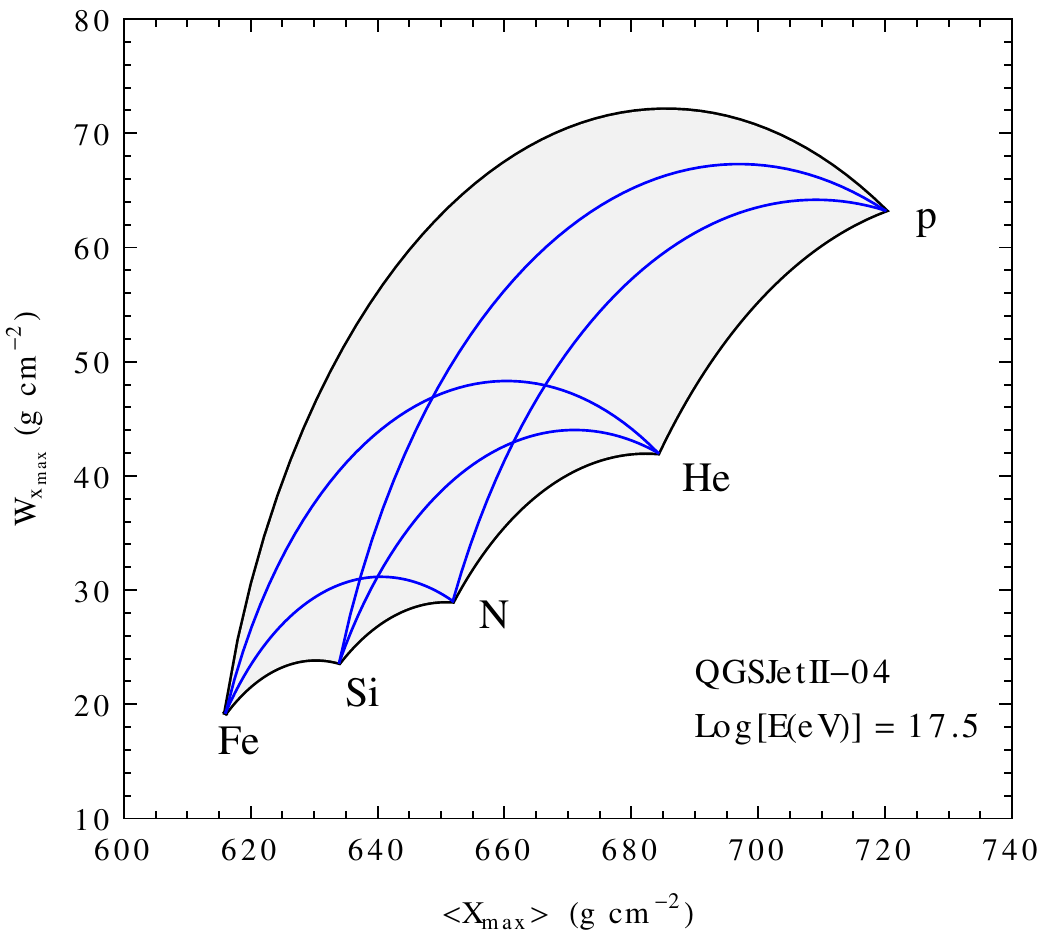}
\end{center}
\caption {\footnotesize
 Allowed region in the plane
 $\{\langle X_{\rm max}\rangle, W \}$ calculated 
 using the hadronic model QGSJetII-04 \cite{Ostapchenko:2013pia} at the energy
 $E= 10^{17.5}$~eV and considering the contributions of 5 nuclei
 ($p$,
 ${}^4$He, 
 ${}^{14}$N, 
 ${}^{28}$Si and
 ${}^{56}$Fe). The lines show points that can be generated by the
 combinations of two nuclei.
\label{fig:region_long0} }
\end{figure}

For any value of the energy, and fixing the model for shower development,
there is an allowed region in the plane $\{X_{\rm max}, W\}$,
that is a set of values that can be obtained for a possible combination of nuclei.
One example of such an allowed region 
(for $E = 10^{17.5}$~eV and using the QGSJetII--04 model)
is shown as a shaded area in Fig.~\ref{fig:region_long0}.
This region has been calculated assuming for simplicity
that only five nuclei 
($p$,
${}^{4}$He, 
${}^{14}$N, 
${}^{28}$Si and
${}^{56}$Fe)
give non--negligible contributions to the CR spectrum.
Because of the poor mass resolution of the observations this is
a good approximation if one interprets each component as 
the sum of contributions of the nuclei in appropriate
mass number intervals. This description of the composition
has been widely used for the study
of high energy cosmic rays.

In the figure one can easily identify
the five points that correspond to
pure compositions, while the curved lines that connect two such points
corresponds to all pairs of values $\{\langle X_{\rm max}\rangle, W\}$ that can be obtained
with compositions formed by two components.
The boundary of the allowed region 
is formed by a subset of these two--component lines.
The upper part of the boundary is the curve for proton--iron combinations,
and the lower part of the boundary
is formed by combinations of two elements that are adjacent in mass:
proton--helium, helium--nitrogen, nitrogen--silicon and silicon--iron.
It should be noted that if the point
$\{\langle X_{\rm max} \rangle, W\}$ is near the boundary of the allowed region
the composition is a two component mixture
and is uniquely determined.

\begin{figure}[bt]
\begin{center}
\includegraphics[width=11.0cm]{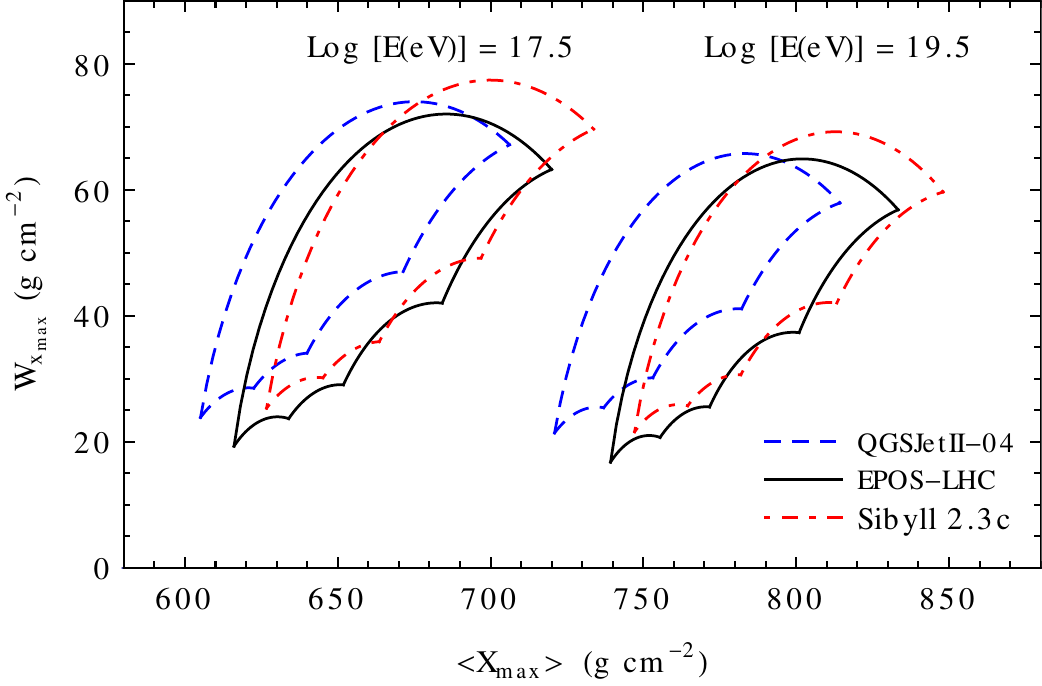}
\end{center}
\caption {\footnotesize
Allowed regions in the plane
 $\{\langle X_{\rm max}\rangle, W \}$ calculated using 
 three hadronic models
 (QGSJetII-04, Epos--LHC and Sibyll 2.3c) for two energies
 $E = 10^{17.5}$ and $10^{19.5}$~eV.
\label{fig:region_long1} }
\end{figure}

The allowed region in the plane $\{\langle X_{\rm max} \rangle, W\}$
changes with energy and is determined by the model.
This is illustrated in Fig.\ref{fig:region_long1} that shows
the allowed region for the three models introduced above,
and for two values of the primary particle energy
($E = 10^{17.5}$~eV and $E = 10^{19.5}$~eV).
Increasing the energy the allowed region moves to higher values
of $\langle X_{\rm max} \rangle$, while the width $W$ changes only slowly.

From Fig.\ref{fig:region_long1} one can see that 
the position of the allowed region is energy
and model dependent, but has a shape that remains
in good approximation constant. 
This suggests to study the evolution with energy of
the CR mass composition introducing rescaled (adimensional) variables:
\begin{equation}
x = \frac{\langle X_{\rm max} \rangle - X_{\rm Fe}}{X_p - X_{\rm Fe}}
\label{eq:xscale}
\end{equation}
and
\begin{equation}
y = \frac{W - W_{\rm Fe}}{W_p - W_{\rm Fe}} ~.
\label{eq:yscale}
\end{equation}
In these expression $\langle X_{\rm max} \rangle$ and $W$ are
obtained from the data, while the other quantities
must be calculated using a model for shower development.

In good approximation a point in the plane
of the rescaled variables $\{x, y\}$ is mapped to the same values
of the mass fractions $\{f_A\}$ independently from the energy,
therefore the trajectory of the point in this plane that describes the 
measurements (for different energies) 
is a good method to visualize the evolution of the CR composition.

\begin{figure}[bt]
\begin{center}

\includegraphics[height=7.5cm]{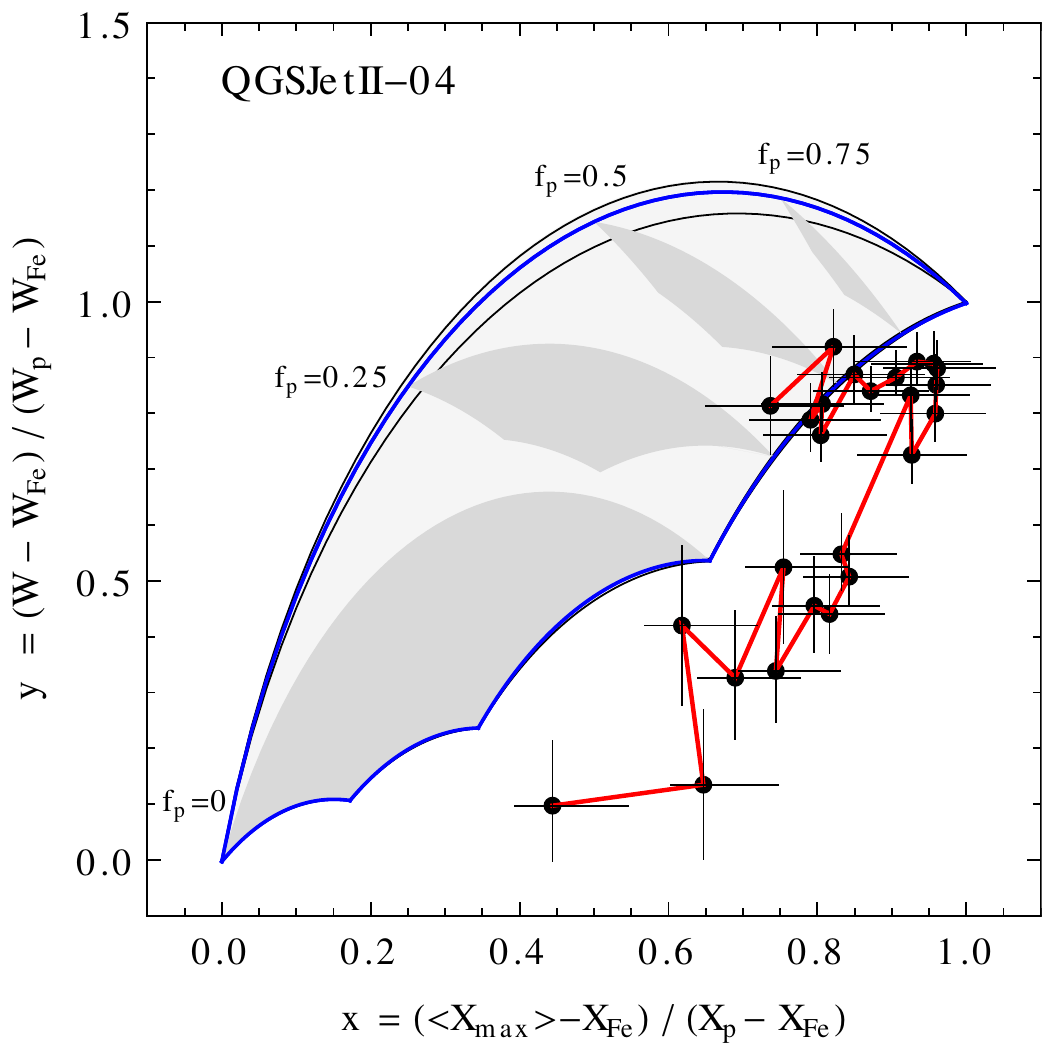}
~\includegraphics[height=7.5cm]{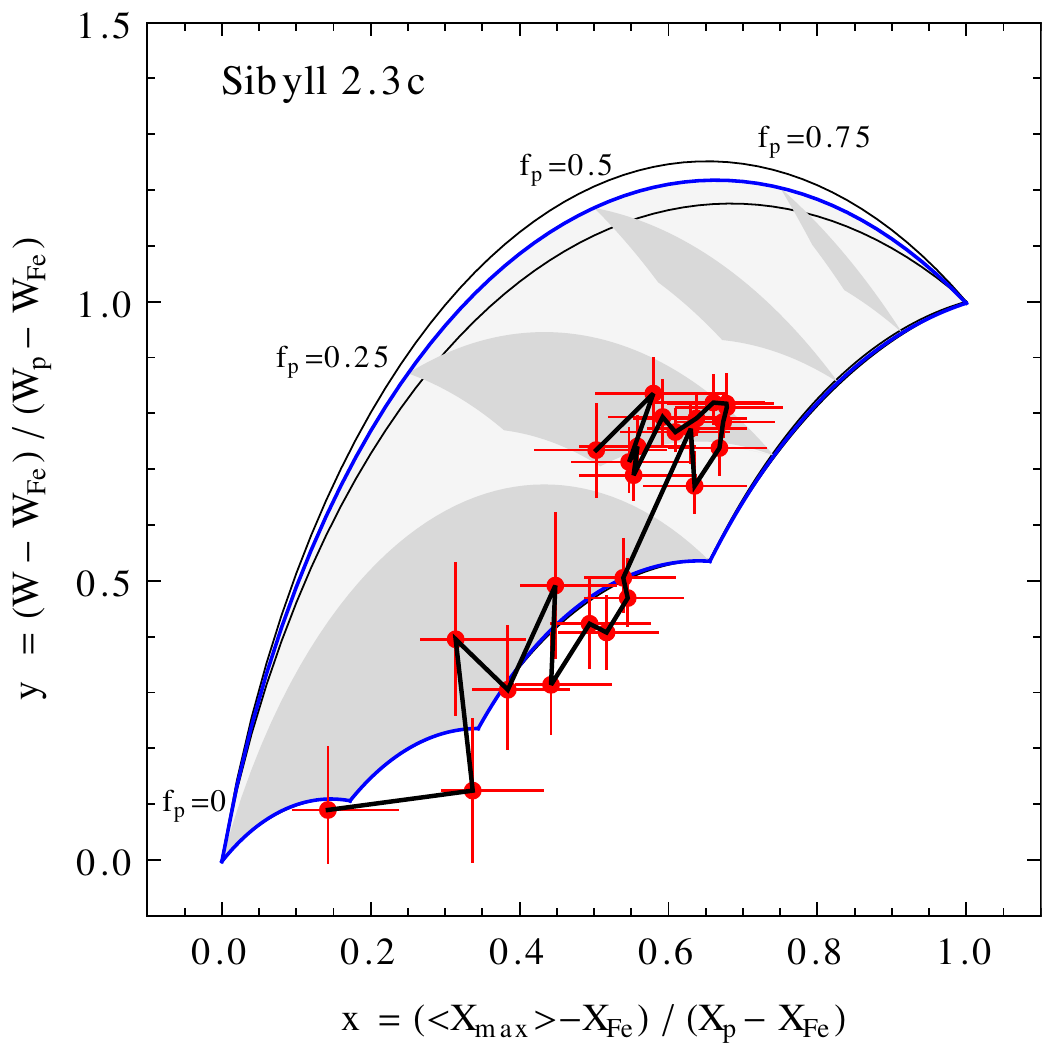}

\vspace{1.1 cm}

\includegraphics[height=7.5cm]{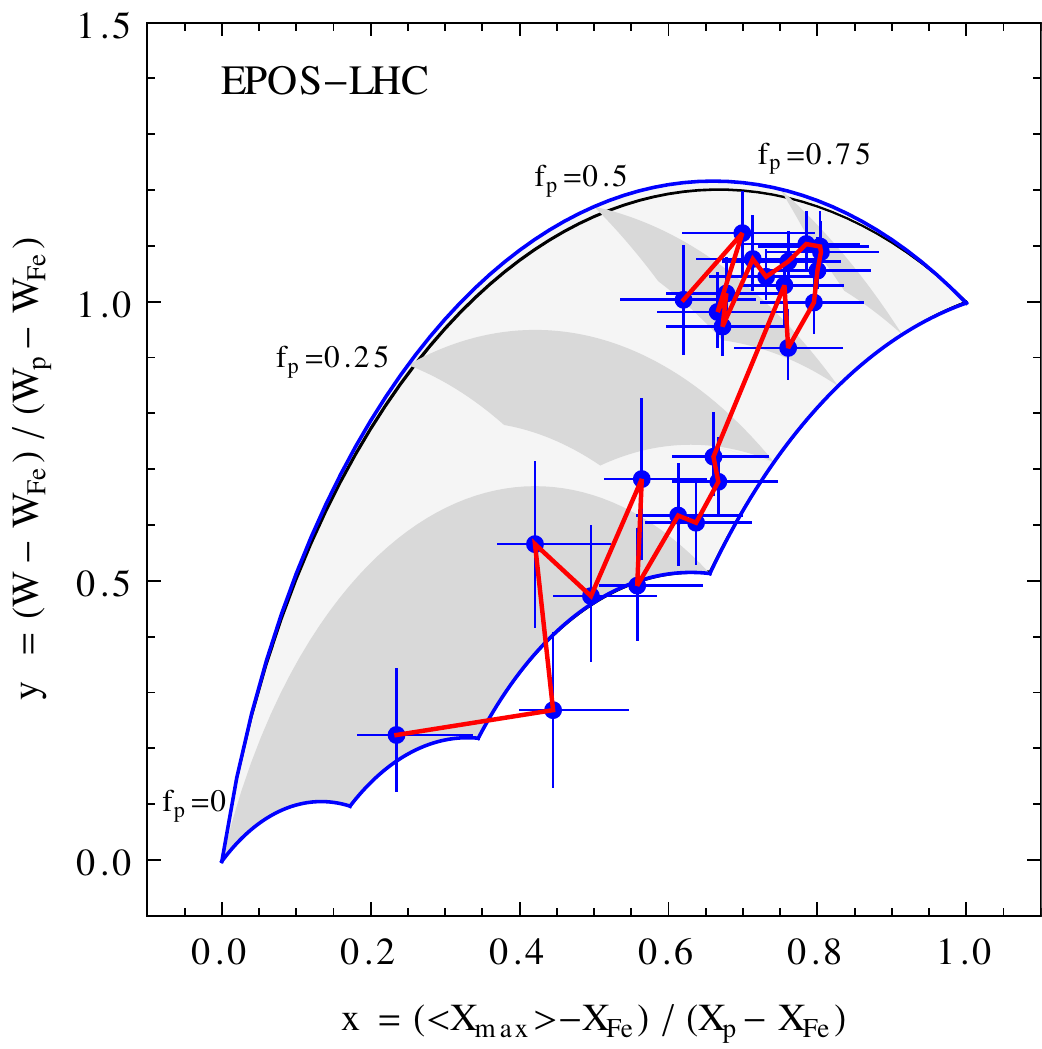}
~\includegraphics[height=7.5cm]{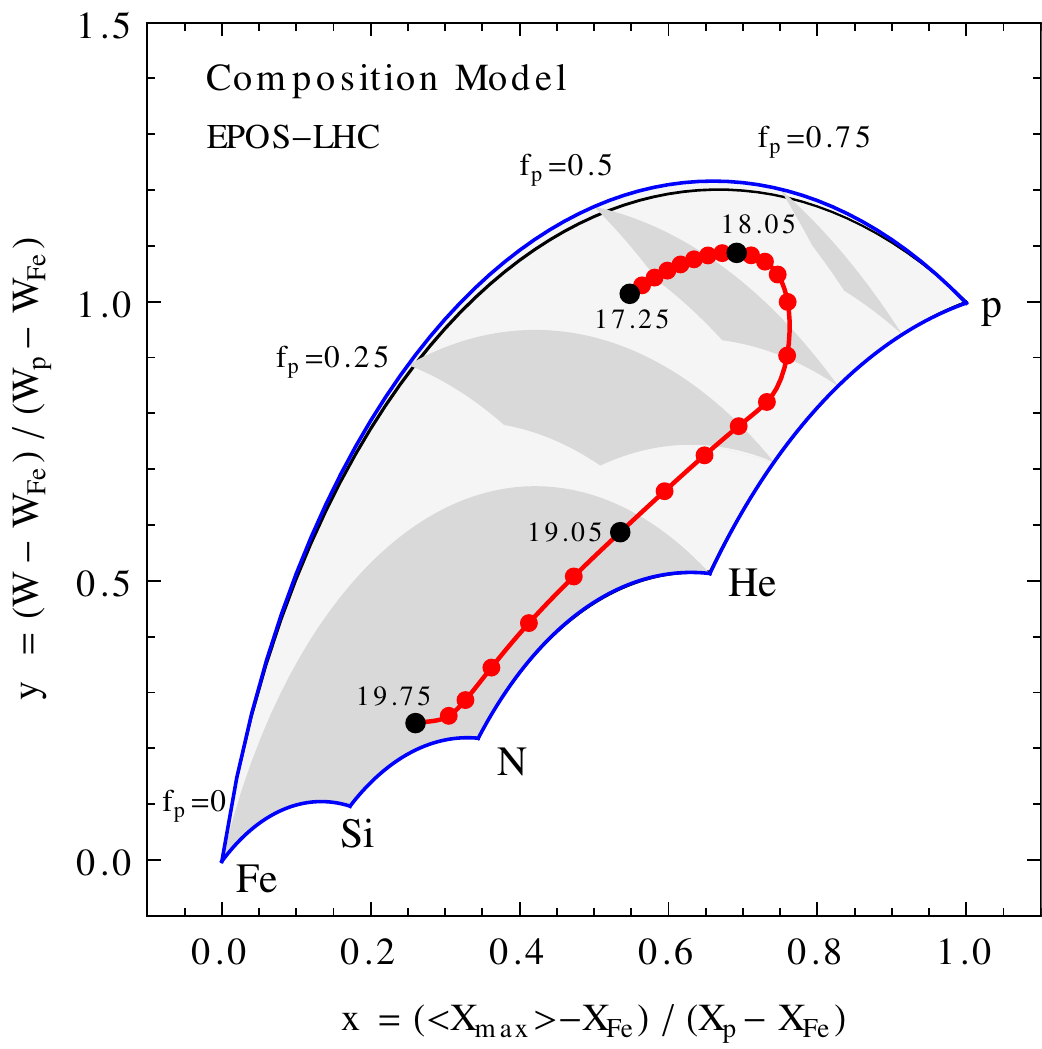}

\end{center}
\caption {\footnotesize
 \label{fig:comp_scheme}
 The Auger measurements of the average and dispersion
 of the depth of maximum distributions (shown in Fig.~\ref{fig:auger_long})
 are represented as points in the
 plane of the rescaled variables $x$ and $y$ [see Eqs.~(\ref{eq:xscale})
 and~(\ref{eq:yscale})].
 Three panels show the results for the three
 hadronic models QGSJetII-04, Sibyll~2.3c and EPOS--LHC.
 The broken line connects data points in adjacent energy intervals, with the
 highest energy point the one with the lowest $x$ and $y$ values.
 The last panel
 shows the trajectory in the space $\{x,y\}$ for the composition model
 discussed in the text (and shown in Fig.~\ref{fig:spectrum_cr_comp})
 calculated using the EPOS-LHC model (representative values of $\log[E ({\rm ev})]$
 are labeled). In all panels the shaded area shows the region of parameters space
 allowed for the combination of five nuclei considered.
 The darker shaded areas indicate the parameter regions that are physically possible
 for a fixed proton fraction (with values
 $f_p = 0.75$,
 $f_p = 0.5$, 
 $f_p = 0.25$ and 
 $f_p = 0$ as marked, a pure proton composition ($f_p =1$) corresponds to the corner
 at the upper right).
}
\end{figure}

This idea is illustrated in Fig.~\ref{fig:comp_scheme}
where the Auger measurements of $\langle X_{\rm max} \rangle$ and $W$ at
energies are shown as points with error bars in the plane $\{x,y\}$
after rescaling the results
[with Eqs.~(\ref{eq:xscale}) and~(\ref{eq:yscale})]
using the theoretical predictions of the three models
(QGSJetII--04, EPOS--LHC and Sibyll~2.3c). 

The first panel in Fig.~\ref{fig:comp_scheme} shows the results
for the QGSJetII--04 model. In this case the points at high energy 
are outside the allowed region indicating that the model is not viable.
The second and third panel show the results using the
Sibyll~2.3c and EPOS--LHC models. Using these models the Auger observations 
can have a consistent interpretation, however the evolution of the
composition indicated by these studies
has some very remarkable and surprising properties, as discussed in
the next section.

In Fig.~\ref{fig:comp_scheme} are also shown the regions in the plane
$\{x, y\}$ plane that correspond to certain representative
values of the proton fraction ($f_p = 1$, 0.75, 0.5, 0.25 and 0) are indicated.
This allows to note that at high energy
the points that describe the measurements move toward small values of the proton fraction.

\section{Interpretation of the Auger depth of maximum measurements}
\label{sec:interpretation}
In this section we will discuss the interpretation of
the fluorescence light observations of Auger,
assuming that the models of shower development used in the Monte Carlo
simulations are correct.

As discussed above, Auger has measured an elongation rate that is larger
than the constant composition prediction below the energy $E^* \approx 10^{18.3}$~eV,
and smaller above. This implies that the CR composition is continuously
changing, first (for $E < E^*$) becoming gradually lighter
and then gradually heavier. 
Below $E^*$ one also observes that
the width of the depth of maximum distribution
is approximately constant, while at higher
energy it decreases monotonically. These observations also play
an important role in determining the evolution of the CR composition.

The energy $E^*$ where one observes these effects
is close (even if not identical) to the energy where
the all--particle spectrum exhibits the sharp hardening
commonly known as the ``ankle''. This suggest to ``identify''
$E^*$ and the ankle energy, assuming that the
spectral feature and the changes in composition have a common origin.
In the following we will first discuss the CR composition 
at $E \approx E^*$ and then its evolution below and above $E^*$.

\subsection {Composition for $E \approx E^*$}
A good determination of the CR composition around the energy $E^*$,
where it is the lightest, is a crucial element
to develop an understanding of very high energy cosmic rays.
Observations of the shape of the tail of
the depth of maximum distribution \cite{auger-sigma-2012}
(that we will discuss in more detail below)
indicate that around this energy the spectrum contains
a large proton component.
The estimate of the fraction of the spectrum formed by protons
is however model dependent.

Inspecting Fig.~\ref{fig:auger_long}
one can see that at $E \simeq E^*$ the QGSJet~II--04 model
predicts for a pure proton composition an $X_{\rm max}$ distribution that
(within errors) has the same average and width of the data.
This model however cannot provide a consistent interpretation of the data
because the measurements of $\langle X_{\rm max} \rangle$ and $W$
at higher energy fall outside the allowed region predicted by the model,
as discussed in the previous section.
The predictions of the EPOS--LHC and Sibyll~2.3c
for the average depth of maximum of a pure proton composition
are larger than the Auger measurement,
and therefore according to these models
the spectrum contains a component of higher mass nuclei.

It is instructive to discuss the case where the spectrum
is formed by only two components:
protons (that account for a fraction $f_p$ of the spectrum),
and nuclei of mass number $A$ (that account for a fraction $1-f_p$).
Using Eq.~(\ref{eq:simple_average}) the average and
width of the depth of maximum distribution at energy $E$ are:
\begin{equation}
\langle X_{\rm max} \rangle = \langle X_p \rangle - (1-f_p) \; D_0 \; \log A~.
\label{eq:xav_prot}
\end{equation}
\begin{equation}
\langle W^2 \rangle = f_p \, W_p^2 + (1-f_p) \, W_A^2 + f_p \, (1-f_p)~D_0^2 \; (\log A)^2
\label{eq:ww_prot}
\end{equation}
(where we have left implicit the energy dependence).
Using a model for the predictions of 
the elongation rate $D_0$, the proton average depth of maximum $\langle X_p \rangle$ and the
widths $W_p$ and $W_A$ of the two distributions, 
Eqs~(\ref{eq:xav_prot}) and~(\ref{eq:ww_prot}) allow to obtain
both the proton fraction $f_p$ and the mass number $A$ of the
second component from the measurements of $\langle X_{\rm max} \rangle$ and $W$.

\begin{figure}[bt]
\begin{center}
\includegraphics[width=14cm]{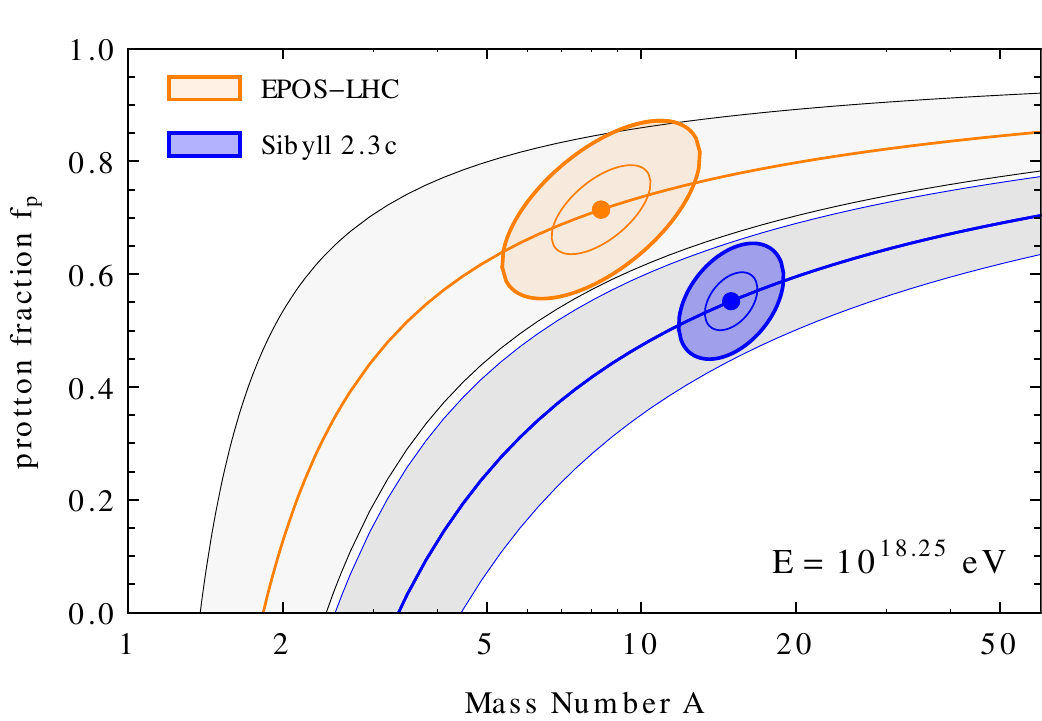}
\end{center}

\caption {\footnotesize
 \label{fig:comp1_18}
 Interpretation of the Auger measurements $\langle X_{\rm max} \rangle$ and $W$
 at energy $E = 10^{18.25}$~eV in terms of a composition formed by protons and a
 second component of mass $A$, using the EPOS--LHC an Sibyll~2.3c models.
 The thick lines show the proton fraction needed to reproduce the
 measured value of $\langle X_{\rm max} \rangle$ as a function of $A$
 (with the shaded area a one sigma uncertainty interval).
 The ellipses show the (one standard deviation) allowed region in the plane $\{A, f_p\}$ calculated
 taking into account the measurement of the width $W$ of the depth of maximum distribution.
}
\end{figure}

\begin{figure}[bt]
\begin{center}
\includegraphics[height=8cm]{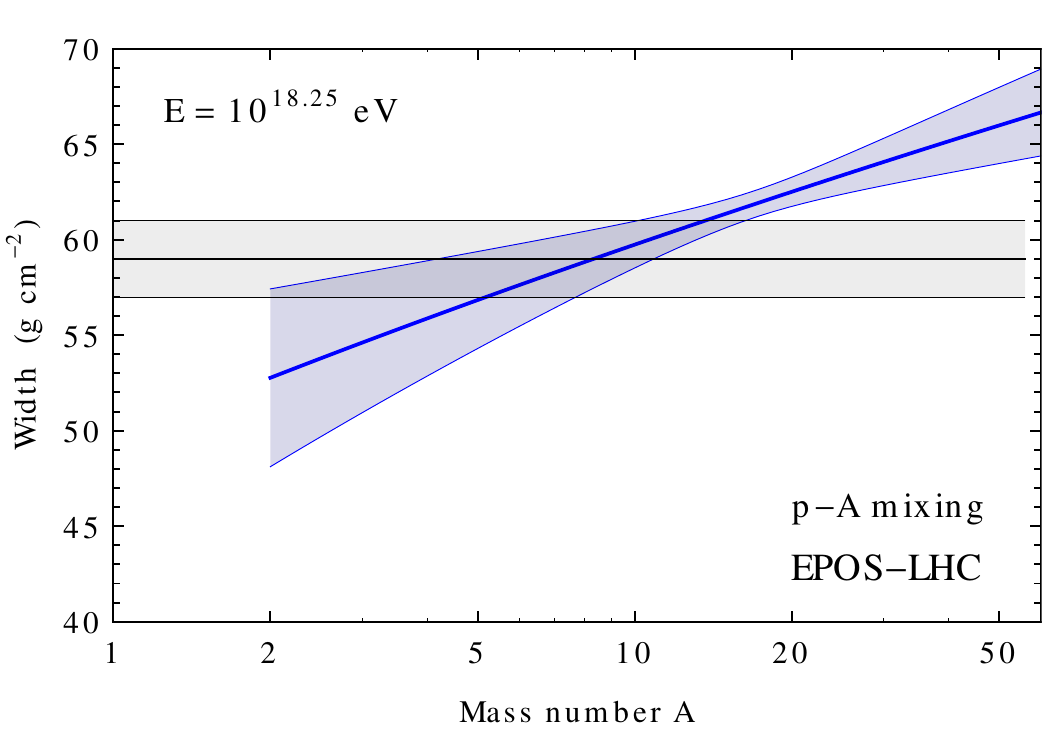}

\vspace{0.4cm}

\includegraphics[height=8cm]{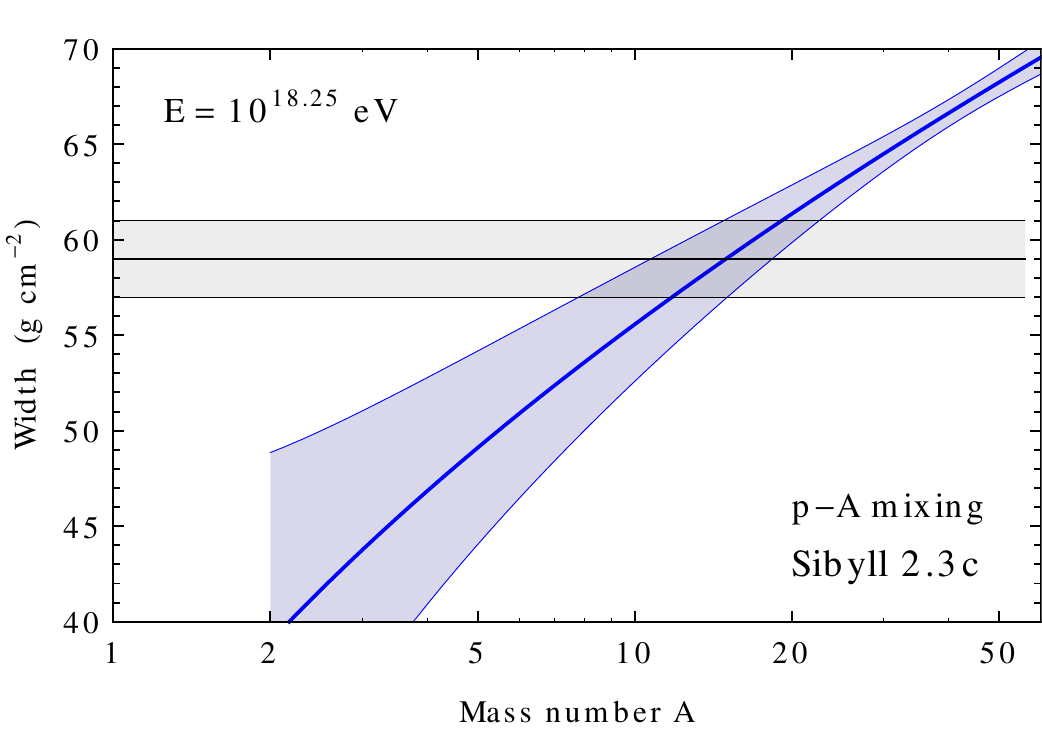}
\end{center}

\caption {\footnotesize
 \label{fig:width_model}
 Width of the depth of maximum distribution predicted
 at the energy $E = 10^{18.25}$~eV for a composition
 formed by protons and nuclei of mass number $A$.
 For each value of $A$ the proton
 fraction $f_p$ is determined by the requirement to reproduce the
 value of $\langle X_{\rm max} \rangle$ obtained by Auger \cite{yushkov-icrc2019}
 (the shaded area is a one sigma uncertainty band).
 The measured value of $W$ (with a 1--$\sigma$ error) is shown as
 the horizontal band.
 The top (bottom) panel uses the EPOS--LHC (Sibyll~2.3c) model.
}
\end{figure}

The results of this exercise at the energy $E = 10^{18.25}$~eV
are shown in Fig.~\ref{fig:comp1_18}.
For EPOS one finds $f_p \simeq 0.71\pm 0.09$ and $A \simeq 8^{+4}_{-3}$,
and for Sibyll $f_p \simeq 0.55\pm 0.07$ and $A \simeq 15^{+5}_{-4}$,
where the (one $\sigma$) errors take into account only uncertainties in the
experimental measurements. There is a positive correlation
between $f_p$ and $A$, because one can obtain the same
average $\langle X_{\rm max}\rangle$ with a smaller nuclear component
of larger mass number.
For both models protons are the most abundant component of the spectrum,
but nuclei are not negligible. 
The fits disfavor compositions where the nuclear component
has a very large $A$, because in this case the predicted
width becomes too large (as illustrated in Fig.~\ref{fig:width_model}),
and the mixing of protons and iron nuclei is not allowed. 
The proton fraction is smaller in Sibyll, because in this model
the showers are more penetrating than in EPOS (by approximately 15~g/cm$^2$),
and therefore a larger contribution
from nuclei is required to lower the average depth of maximum
and obtain agreement with the data. 

\subsection {Composition below the ``ankle''}
The elongation rate measured by Auger
in the energy range $10^{17.25}$--$10^{18.25}$~eV
has been fitted \cite{yushkov-icrc2019}
with a constant value $D_{\rm data} \simeq 77\pm 2$~g/(cm$^2$~decade),
that is significantly larger than the model predictions for an energy independent composition
that are of order $D_0 \simeq 56$--61 (same units).
Using Eq.~(\ref{eq:simple_average}) one finds that the composition if changing with energy
with the average $\log A$ that decreases linearly with $\log E$: 
\begin{equation}
\frac{ d \langle \log A\rangle}{d \log E} \simeq
 - \left ( \frac{D_{\rm data}}{D_{0}} -1 \right ) \simeq -0.3 \pm 0.1~.
\label{eq:loga_evolution}
\end{equation}

The simplest interpretation for this change in composition
is to assume that the CR spectrum in this energy range is formed by two
components: $\phi_\ell (E)$ and $\phi_h (E)$ one ``light'' and one ``heavy''
(with average logarithm of mass number $\log A_\ell$ and $\log A_h$)
that have different spectral shapes, with the light component being harder.

The fraction $f_\ell (E)$ of the light component at the energy $E$ is then:
\begin{equation}
 f_\ell (E) =
 \frac{1}{\Delta \log A} ~
 \left [ \log A_h -
 \frac{\langle X_p (E_0) \rangle - \langle X_{\rm max} (E_0) \rangle} {D_0} \right ] +
 \frac{1}{\Delta \log A}
 ~ \left ( \frac{D}{D_0} -1 \right ) ~\log \frac{E}{E_0}
\label{eq:f_ell}
\end{equation}
(where $E_0$ is an arbitrary reference energy and $\Delta \log A = \log A_h - \log A_\ell$).
According to this equation the fraction $f_\ell (E)$ grows linearly with $\log E$
with a slope $\propto (\Delta \log A)^{-1}$:
\begin{equation}
 \frac{df_\ell}{d\log E}
 \simeq \left ( \frac{D_{\rm data}}{D_0} -1 \right ) ~\frac{1}{\Delta \log A}
 \simeq (0.17 \pm 0.06) ~\frac{\log 56}{\Delta \log A}~.
\end{equation}
A larger value of $\Delta \log A$ 
corresponds to a more slow variation of the composition.
Note that also the constant term in Eq.~(\ref{eq:f_ell}) has the same dependence
on the mass composition of two components $\propto (\Delta \log A)^{-1}$.

In a two--component model, the fraction $f_\ell (E)$ also determines the width of the
$X_{\rm max}$ distribution:
\begin{equation}
 W^2(E) = W_p^2 \, f_\ell (E) + W_A^2 \, [1- f_\ell (E)] +
 f_\ell (E)~[1-f_\ell (E)]\; D_0^2 ~\left (\Delta \log A \right )^2 ~.
\label{eq:width1}
\end{equation}
The data show that in the energy range $10^{17.2}$--$10^{18.3}$~eV the width $W$
is approximately constant with a value of order 60~g/cm$^2$. This is an important
constraint of the possible masses of the two components, that disfavors
$\Delta \log A$ too large.

The fraction $f_\ell$ can only take values in the interval $[0,1]$,
therefore an energy dependence linear in $\log E$ can only be valid in
a limited range. It is therefore interesting to discuss a model for the
energy dependence of the spectral components that can be extended to a a broader range.
A simple scenario is one where the two components have both power--law form,
and can be written as:
\begin{equation}
 \phi_{h (\ell)} (E) = \phi^\dagger ~\left (
 \frac{E}{E^\dagger} \right )^{-\overline{\gamma} \mp \Delta \gamma/2} ~.
\end{equation}
where $E^\dagger$ is a ``crossing energy'' where the
two components are equal (with value $\phi^\dagger$).
In this model the spectral index and the elongation rate (for the all--particle flux)
are both energy dependent: 
\begin{equation}
 \gamma(E) = \overline{\gamma} - \frac{\Delta \gamma}{2}
 ~\tanh \left ( \frac{\Delta \gamma}{2} ~\ln \frac{E}{E^\dagger} \right ) ~,
\label{eq:gamma_energy}
\end{equation}
\begin{equation}
 D(E) = D_{0} ~\left [1 + \frac{\Delta \ln A ~\Delta \gamma}{4}
 ~\cosh^{-2} \left (\frac{\Delta \gamma}{2}
 \; \ln \frac{E}{E^\dagger} \right ) \right ]
\label{eq:elon_evol}
\end{equation}
Eq.~(\ref{eq:gamma_energy}) states that the spectral index has value
$\overline{\gamma} \pm \Delta \gamma/2$ for
$E \ll E^\dagger$ ($E \gg E^\dagger$), changing gradually around the crossing energy,
while Eq.~(\ref{eq:elon_evol}) predicts that the elongation rate is equal to
$D_0$ for energies
much higher and much lower than $E^\dagger$,
and takes a larger value in the intermediate region.

The observations of Auger in the energy range considered \cite{verzi:2019u}
are well described by a simple power law spectrum 
with exponent $\gamma \simeq 3.27 \pm 0.05$, and a constant elongation rate.
The model with two components of power---law form can be made consistent
(taking into account measurement errors) with these observations, because
for $E$ in an energy interval around the crossing energy $E^\dagger$
determined by the condition:
\begin{equation}
 \left | \frac{\Delta \gamma}{2} \ln \frac{E}{E^\dagger} \right | \lesssim \frac{1}{2}
\label{eq:e-interval}
\end{equation}
the arguments of the functions $\tanh$ and $\cosh$ containing the energy dependence
of $\gamma(E)$ and $D(E)$ in Eqs.~(\ref{eq:gamma_energy})
and~(\ref{eq:elon_evol}), are sufficiently small, so that it is possible to
substitute $\tanh x \to 0$, $\cosh x \to 1$, so that 
both quantities can be considered constant with values:
\begin{equation}
\gamma \simeq \overline{\gamma}
\label{eq:gamma_av}
\end{equation}
and
\begin{equation}
 D \simeq D_{0} ~\left [1 + \frac{\Delta \ln A ~\Delta \gamma}{4} \right ] ~.
\end{equation}
The last two equations determine the spectral indices of the two components,
that have average equal to the slope of the all--particle flux, and difference:
\begin{equation}
 \Delta \gamma \simeq 4 \; \left ( \frac{D_{\rm data}}{D_0} -1 \right ) ~
 \frac{1}{\Delta \ln A} \approx 0.3 ~\frac{\log 56}{\Delta \log A}~.
\end{equation}
The difference in spectral index for the two components depends on their masses:
$\Delta \gamma \propto (\Delta \log A)^{-1}$,
and is large when the mass numbers of the two components are close.
A too small $\Delta \log A$ is however not consistent with data
because it corresponds [see Eq.(\ref{eq:e-interval})]
to a too short energy interval where the spectral index and elongation rate
can be considered as constant.
A too large $ \Delta \log A$ is also not viable, because it corresponds to a
width $W$ larger than the measurement. 

In conclusions, the observations of Auger in the sub--ankle region ($E \lesssim 10^{18.3}$~eV),
interpreted with the current models, require
a composition that becomes gradually lighter, with an average
$\langle\log A\rangle$ that changes by a (modest but significant)
$0.3 \pm 0.1$ in the decade between 10$^{17.3}$ to $10^{18.3}$~eV.
This change of composition can be described with the mixing of
protons with nuclei of intermediate mass (with proton--iron mixing disfavored).
A model where the change in composition is due to the combination of
two components of power law form is viable, and results in
spectral indices that differ by $\Delta \gamma \approx 0.35$--0.55,
for compositions with $\Delta \log A \approx 0.7$--1.4.

An explicit example of such a two--components scenario,
constructed on the basis of the EPOS--LHC model,
is shown in Fig.~\ref{fig:spectrum_cr_comp}.
In this model the sub--ankle spectrum is formed by a proton component
and a second component of nitrogen and
silicon nuclei with equal weight. The average spectral indices of the components 
$\overline{\gamma} = 3.27$ is equal to the slope of the all--particle flux, 
the difference in spectral index
is $\Delta \gamma = 0.49$, and the crossing energy is $E^\dagger = 0.19$~EeV.

\subsection{Composition above the ``ankle''}
For $E \gtrsim 10^{18.3}$~eV the elongation rate
measured by Auger is significantly smaller than the constant composition
predictions, indicating that the composition is rapidly evolving toward
a heavier mixture.
In this energy range however the evolution of the composition
cannot be described as the simple combination of two components.
This is the consequence of the energy dependence of the width $W$
that decreases rapidly and monotonically
from $W \approx 60$~g/cm$^2$ to $\sim 30$~g/cm$^2$.

A qualitative understanding of the evolution of the CR composition
can be visualized inspecting Fig.~\ref{fig:comp_scheme} that
shows the rescaled measurements of $\langle X_{\rm max} \rangle$ and $W$
for different values of the energy.
For the QGSJET model the points that represent the measurements
fall outside the allowed region, indicating that the predictions cannot be correct.
For the EPOS ans Sibyll model the points are
inside the allowed region, but close to the lower boundary.
Points on this boundary fully identify the
composition, and correspond to the mixing 
of two elements that are adjacent in mass number.

Using the EPOS--LHC model, for $E \gtrsim 10^{18.5}$~eV,
the composition becomes very rich in helium,
then evolves to a combination of helium and nitrogen, with the nitrogen fraction
that grows with $E$, and at the highest energy
there are indications that silicon begins to mix with nitrogen.

Using the Sibyll~2.3c model, where
the showers are on average approximately 15~g/cm$^2$ deeper,
the data are interpreted with an heavier
composition, but with an evolution with energy that has same qualitative features.
Above the ankle the composition becomes first a mixture
of helium and nitrogen, then of nitrogen and silicon, and finally there is a hint of
the appearance of iron at the highest energies.

The very rapid evolution with energy of the composition
requires that the spectra of the individual elements
are curved, so that they can give a contribution that 
first increases rapidly, and then rapidly disappears.
This requirements can be satisfied in a rather simple model,
that has been presented by Auger
\cite{Aab:2020rhr,Castellina:2019huz}
as the most natural
framework to interpret the data.
In the model
the cosmic rays spectra have a rigidity dependent shape of form:
\begin{equation}
 \phi_A (E) \simeq K_A ~\left (\frac{E}{E_0} \right )^{-\gamma_0} ~
 f_{\rm cut} \left (\frac{E}{Z \,E_{\rm cut}} \right)
\label{eq:spectrum_peters}
\end{equation}
where $A$ and $Z$ are the mass number and electric charge of the nucleus,
and $f_{\rm cut}(x)$ is a cutoff function
that is unity for $x \lesssim 1$ and falls rapidly to zero for $x > 1$.
The spectra of each element has then a cutoff at an energy that
increases linearly with $Z$.
Using this form, the relative contribution of
nuclei of charge $Z$ can become dominant in a narrow energy range
before its own cutoff, but above the cutoff of nuclei with smaller $Z$.
In this scenario it is then possible to choose the parameters
of the spectra so that the all--particle
spectrum is dominated by protons for $E \lesssim E_{\rm cut}$,
then by helium for $E_{\rm cut} \lesssim E \lesssim 2\; E_{\rm cut}$,
and by more and more massive nuclei as the energy increases.

\begin{figure}[bt]
\begin{center}
\includegraphics[width=16.0cm]{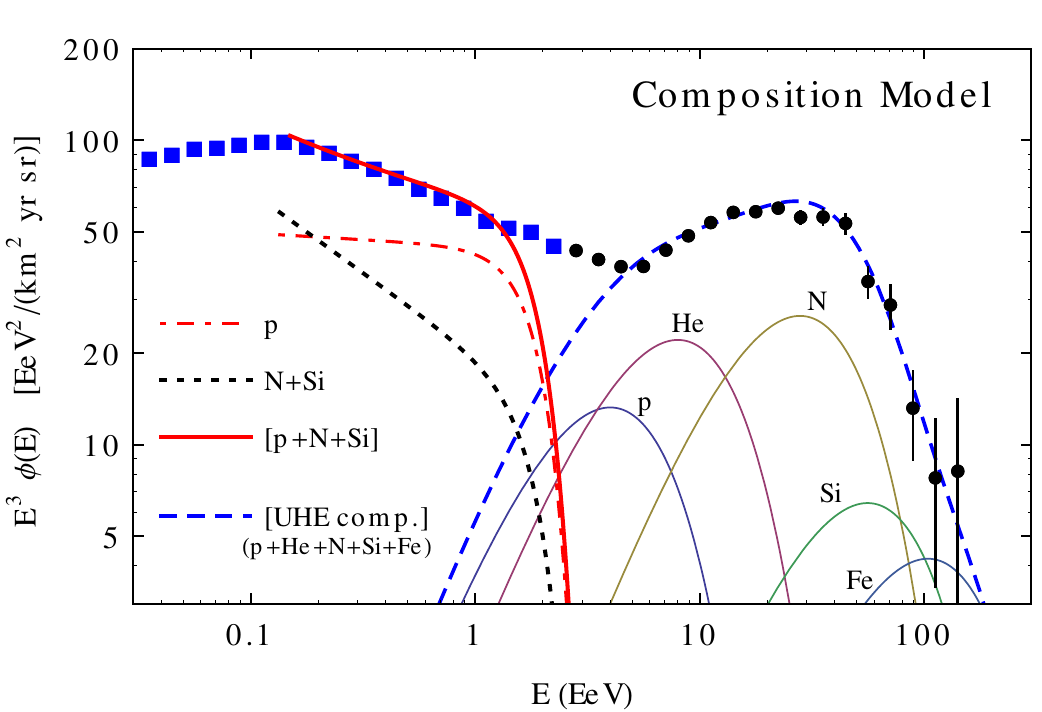}
\end{center}
\caption {\footnotesize
 Model of the CR energy spectrum and composition constructed to
 reproduce the Auger data.
 The spectrum below the ankle (shown as a thick solid line)
 is formed by two component, one of protons and the
 other of nitrogen and silicon (with equal abundances) that have both power law form
 with superexponential cutoffs. The spectrum above the ankle (thick dashed line)
 is formed by the contributions of five nuclei (protons, helium, nitrogen, silicon and iron)
 that have a hard power law spectra (with the same slope), with rigidity dependent
 cutoffs (see main text for more details).
 The data points are from Auger \cite{Aab:2020gxe,verzi:2019u}.
\label{fig:spectrum_cr_comp} }
\end{figure}

An example of this scenario is shown in Fig.~\ref{fig:spectrum_cr_comp}
where the cosmic ray flux above the ankle is fitted as the
combination of five nuclei with spectra
of the form of Eq.~(\ref{eq:spectrum_peters}) using for the cutoff
a simple exponential form: $f_{\rm cut}(x) = e^{-x}$. The parameters
of the fit are: $\gamma_0 = 1$, $E_{\rm cut} = 1.8 \times 10^{18}$~eV,
and relative mass fractions 
$p:{\rm He}:{\rm N}:{\rm Si} :{ \rm Fe} = 1 : 0.5:0.05:0.003:0.0004$.
It should be noted that these parameters describe the
CR composition at the Earth. If the CR particles in this energy range are
extragalactic, the composition is modified
during propagation because of photodisintegration reactions, and
a model of the injection is required to infer the composition at the source.

The high energy component described above
accounts for the total of the all--particle flux for energies $E \gtrsim 10^{18.5}$~eV.
At lower energy, is contributes only a very small fraction
of the total flux, because of its very flat spectrum.
On the other hand, the sub--ankle component
(modeled as a power law in the discussion above)
cannot continue featureless at higher energy,
because in this case it would contribute a non negligible fraction to the total,
spoiling the results on composition discussed above.
It is therefore necessary to introduce 
a sharp cutoff for the sub-ankle components (as shown in Fig.~\ref{fig:spectrum_cr_comp}).

\subsection{The proton fraction}
\begin{figure}[bt]
\begin{center}
\includegraphics[width=14cm]{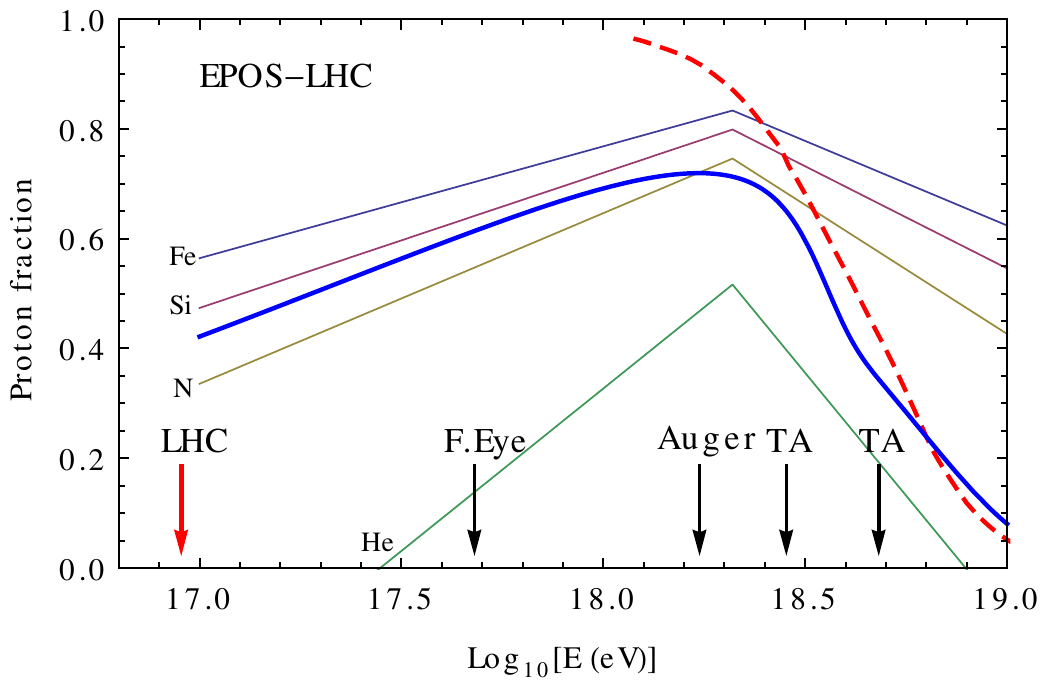}
\end{center}
\caption {\footnotesize
 \label{fig:proton_fraction}
 Fraction of protons in the CR flux as a function of energy.
 The thin lines are estimates
 of the proton fraction obtained from
 the measurements of the average depth of maximum obtained by
 Auger \cite{yushkov-icrc2019} comparing with the predictions of 
 the EPOS--LHC model and assuming that the composition is formed by
 the combination of protons and one nuclear component 
 (helium, nitrogen, silicon and iron) [see Eq.~(\ref{eq:f_ell})].
 The thick solid line is the proton fraction for the model
 discussed in the text (and shown in Fig.~\ref{fig:spectrum_cr_comp}).
 The dashed line is the proton fraction in the very high energy component
 discussed by the Auger collaboration \cite{Aab:2020rhr,Castellina:2019huz}.
 The Auger model does not include the ``sub--ankle'' component, but is in good agreement
 with the ``super--ankle'' component discussed in this paper.
}
\end{figure}

An important result obtained interpreting the Auger results
with the current models is the energy dependence of the proton
fraction. The form of this energy dependence is illustrated in
Fig.~\ref{fig:proton_fraction}, using as model EPOS--LHC.
Modeling the CR flux as the combination of
two components of protons and nuclei of mass number $A$,
the comparison of the model with the measurement of $\langle X_{\rm max} \rangle$
is sufficient to determine the proton fraction $f_p$
(that can be obtained using Eq.~(\ref{eq:f_ell}) with $A_\ell =1$).
Curves of $f_p (E)$ obtained in this way
for $A = 4$, 14, 28 and 56 are shown in Fig.~\ref{fig:proton_fraction}.
As already discussed, the measurement of the elongation rate
(with a break at $E \simeq 10^{18.3}$~eV), requires that the
proton fraction is energy dependent. One can also see
that for a smaller mass number $A$, the proton fraction
$f_p$ is smaller and changes more rapidly with energy.

The estimate of the proton fraction requires a model for the mass distribution
of cosmic rays at different energies.
In this section we have constructed an example of such a model for the composition,
where the spectrum is formed by the sum of sub--ankle and super--ankle components,
with the first one formed by two subcomponents (protons and nitrogen plus silicon)
of power--law form with a sharp (super--exponential cutoff),
while the second is formed by subcomponents that have the same
hard spectrum with rigidity dependent exponential cutoffs.
The proton fraction for this model is shown
(as a thick solid line) in Fig.~\ref{fig:proton_fraction}. The $p$--fraction
grows first (below the ankle) rather slowly, linearly in $\log E$ with a slope 0.47/decade,
it reaches a maximum value around $E \simeq 10^{18.3}$~eV, where both components
(that at this energy are both rich in protons)
give significant contributions,
and then (above the ankle) falls very rapidly after the cutoff of the
high energy proton component.
In the figure is also shown (as a dashed line)
the proton fraction estimated by Auger
\cite{Aab:2020rhr,Castellina:2019huz} for the super--ankle component,
that is in reasonable good agreement with the results obtained here.
The rapid disappearance of protons at high energy is required in these models
to account for the small elongation rate together with the narrowing
width of the depth of maximum distribution.

In the bottom part of Fig.~\ref{fig:proton_fraction} the arrows show the energies
where different cosmic ray experiments
\cite{Baltrusaitis:1984ka,auger-sigma-2012,Abbasi:2015fdr,Abbasi:2020chd},
using the fluorescence light observations,
have obtained measurements of the proton--air cross section.
These measurements cover an energy range where,
according to the composition studies we have discussed above,
the fraction of protons in the CR flux is not constant.
In particular, it is puzzling to note that the two measurements obtained by Telescope Array
\cite{Abbasi:2015fdr,Abbasi:2020chd} are performed at energies where Auger indicates
that the proton component begins to be significantly suppressed.

It is natural to investigate the possibility to use the methods
that have allowed the measurement of the $p$--air cross section, and have therefore
identified a proton component, to try to obtain a quantitative estimate
of its size.
Such an independent measurement of the proton fraction
would be a very significant test of the soundness of the CR composition studies,
and of the validity of the shower models used for
the interpretation of the data.

\subsection{Difficulties for astrophysical models}
The evolution of the cosmic ray spectra that emerges from the Auger
fluorescence light observations, interpreted with current models for
shower development, has some remarkable and unexpected properties,
that had not been predicted by any theory, with
very important implications for our understanding of the
high energy sources. 

\begin{enumerate}
\item The spectra of the
 CR component that dominate at the highest energy, below
 the cutoff are remarkably hard.
The best fit value obtained by the Auger collaboration in
\cite{Aab:2016zth} is $\gamma_0 = 1.22$, in the present work
we have used $\gamma_0 = 1$, that also provides a good description
of the data, and that can have some speculative theoretical motivations. 
This very hard shape is not observed directly, 
because this component becomes visible 
only in the energy range where the spectra 
have sharp cutoffs, however its form can be inferred by the
very fast change in the average mass of the CR particles
with increasing energy.
Such a hard spectral shape is very different than what is observed
for cosmic rays at lower energy, and suggests that in this
range a new, different acceleration mechanism is operating. 

\item A key ingredients of the scenario is the existence of sharp rigidity
 dependent cutoffs in the spectra.
 The need for a well defined cutoff for each component is
 again required by the observation of the very rapid change
 in mass composition with energy.
 What is surprising here is not
 the fact that the cutoffs depend on rigidity, because such
 dependence is in fact predicted by most acceleration models,
 but the ``sharpness'' of these spectral features.
 It is essentially certain, because of limits on the CR anisotropy,
 that several sources contribute to the generation of the highest
 cosmic rays. A well defined cutoff therefore
 implies that different sources generate
 (for each element) spectra with the same maximum energy.
 This is a very important constraint for the properties (such as
 geometrical size and magnetic field) of the sources,
 that calls for an explanation.

\item The shape of the all--particle energy spectrum above the ankle
 is quite smooth, (even is the Auger collaboration has recently presented 
 evidence \cite{Aab:2020gxe} that the energy distribution
 presents a break and cannot be fitted with a simple power law).
 The all--particle
 spectrum is however obtained summing components that all have 
 shapes that are rapidly changing with energy, and the smoothness of
 the observed spectrum emerges
 because the relative sizes of the components are sufficiently ``fine--tuned''.
 The fit performed here (and by Auger \cite{Aab:2020rhr,Castellina:2019huz})
 is purely phenomenological, and considers the fraction of different elements
 in the spectrum as free parameters.
 This yields a good fit to the all--particle
 spectrum, but an important question is 
 what the results imply for the mechanism and environment of particle acceleration.
 In \cite{Aab:2016zth} the Auger collaboration, after modelling the
 propagation effects, has estimated the mass fractions of the spectra
 at the source as
 $f_p : f_{\rm He} : f_{\rm N} : f_{\rm Si} \approx 0.06 : 0.46 : 0.37 : 0.09$ 
 (with a negligible contribution with iron). It is far from easy to construct
 a realistic astrophysical model to generate this composition of 
 accelerated particles.

\item The Auger collaboration until now has only discussed a model
 for the cosmic ray spectra above the ankle.
 This appears to be a very significant limitation because the
 study of the composition in the ``transition region'', where the
 two components are of similar size, can give very valuable information.
 A more complete model of the spectra must clearly include a discussion of
 both the sub--ankle and super--ankle ranges.

 Below the ankle the all--particle CR spectrum can be described by a simple 
 power--law form with a spectral index of order $\gamma \simeq 3.27$. However,
 according to the current models, the composition is this energy range 
 is changing, and this requires that the spectra of different elements
 have different shapes, and the simple power--law of the
 all--particle spectrum emerges only as the sum of these sub--components.
 The ``standard scenario'' where cosmic rays in this energy range are formed
 by an iron--rich component (perhaps of Galactic origin) and a light
 extragalactic one do not give a good description of these observations, and
 one remains with the difficult task to construct a viable astrophysical model.
 
\item A very important result is also the fact that the sub--ankle component
 is required to have a sharp cutoff around the ankle energy.
 This is because if this component of the spectrum
 (that is observed to be rich in protons) continues without a break
 to higher energy, it would form a large fraction of the all--particle
 spectrum, in conflict with the results on composition at very high energy.
 If the sub--ankle spectrum is formed by more than one component,
 as it is suggested by the energy dependence of composition,
 all important components must have cutoffs, at approximately the same energy.
 To avoid excessive fine--tuning, this suggests the need to
 construct a model of CR acceleration
 where the sub--ankle and super--ankle components are related.
 Ideas for such a common origin have been proposed
 \cite{Unger:2015laa}, but the construction of a
 model that describes the CR spectrum and composition across the ankle
 remains a very difficult task.

\end{enumerate}

The list of ``difficulties'' presented above show how interesting
are the results obtained by Auger that, far from being ``disappointing'',
are in fact quite extraordinary.
These considerations suggest however that it is very desirable to
perform additional studies that have the potential to give independent support
to the validity of the Monte Carlo simulations that play an essential role
in the interpretation of the data.
An interesting possibility is the study of the proton component,
as discussed in the next section.

\section{Measurement of proton-air cross section}
\label{sec:sigma}
The shape of the depth of maximum distribution has also been used to measure
the proton--air cross section at very high energy.
The fundamental idea behind this measurement method was first developed by the Fly's Eye 
collaboration \cite{Baltrusaitis:1984ka} and used 
to obtain an estimate of the cross section for a proton
laboratory energy $E \simeq 5 \times 10^{17}$~eV, that corresponds 
to a c.m. energy for a p--nucleon interaction $\sqrt{s} \simeq 30$~TeV.
More recently estimates of $\sigma_{p{\rm Air}}$ have been obtained
by the Auger Collaboration at $\sqrt{s} = 57$~TeV \cite{auger-sigma-2012} 
and by the Telescope Array Collaboration at $\sqrt{s} = 95^{+5}_{-8}$~TeV
\cite{Abbasi:2015fdr} and $\sqrt{s} = 73$~TeV \cite{Abbasi:2020chd}.
These papers
argue that the measurement of the proton cross section is in good approximation
model independent, and is also only weakly dependent on the
composition of the cosmic ray flux, as long as the fraction of protons
is not too small.

The air shower measurement of the $p$--air cross section
are based on the study of the longitudinal development of the showers.
An ideal measurement of the profile would allow
to observe the point $X_0$ where a primary proton
undergoes its first inelastic interaction. The distribution of $X_0$,
for a fixed value of the energy, is a simple exponential:
\begin{equation}
F_0 (X_0) = \frac{1}{\lambda_p} ~e^{-X_0/ \;\lambda_p} 
\label{eq:fx0}
\end{equation}
completely determined by the $p$--air interaction length $\lambda_p$, that can be
calculated from a combination of the interaction cross sections of protons 
with the different nuclei that compose the atmosphere:
\begin{equation}
 \frac{1}{ \lambda_p }
 = \frac{\sum_A ~p_A ~\sigma_{pA}}{\sum p_A ~m_A}
 = \frac{\sigma_{p{\rm Air}}}{\langle m \rangle}
\end{equation}
where the summation runs over all nuclei in air,
$m_A$ is the mass and $p_A$ the relative abundance for
nuclei of type $A$, and the cross section $\sigma_{pA}$ is the so called
inelastic production cross section for $p$--nucleus collisions
obtained substracting from the total cross section
the elastic and quasi--elastic (target--fragmentation) contributions,
that are essentially invisible in the development of a shower.

The existing detectors do not
have the resolution to observe the first interaction point,
and therefore cannot simply measure the $X_0$ distribution and extract 
$\lambda_p$ from its shape.
The idea introduced by the Fly's Eye collaboration is that fluctuations
in $X_0$ are the dominant effect in the fluctuations of the depth of maximum
$X_{\rm max}$ for deeply penetrating showers, so that 
the distribution $F(X_{\rm max})$ of depth of maximum of the showers 
takes asymptotically (for large values of $X_{\rm max}$) 
the exponential form $F(X_{\rm max}) \propto e^{-X_{\rm max}/\Lambda}$
with $\Lambda \approx \lambda_p$. 
The measurement of this asymptotic shape and of its slope
allows then a determination of the $p$--air interaction length.

To illustrate this point more quantitatively one can
note that decomposing the depth of maximum into the sum
\begin{equation}
X_{\rm max} = X_0 + Y \;,
\label{eq:y_def}
\end{equation}
where $X_0$ is the position of first interaction point
and $Y$ the maximum of the shower development measured from this origin,
the $X_{\rm max}$ distribution can be written as the convolution:
\begin{eqnarray}
 F(X_{\rm max})
 & = &\int_0^\infty dY ~\int_0^\infty dX_0 ~G(Y) ~\frac{e^{-X_0/\lambda_p}}{\lambda_p}
 ~\delta[X_{\rm max} - (X_0 + Y)] \nonumber \\
& ~& \nonumber \\
 & = & \frac{e^{-X_{\rm max}/\lambda_p}}{\lambda_p} 
\; \left [\int_0^{X_{\rm max}}
 dY ~G(Y) ~e^{Y/\lambda_p} \right ] ~
\label{eq:convolution}
\end{eqnarray}
where we have used the fact that the distribution of $X_0$ is the simple exponential
given in Eq.~(\ref{eq:fx0}) and $G(Y)$ is the distribution of $Y$,
that is determined by the properties of the hadronic interactions
and therefore model dependent.
Inspecting Eq.~(\ref{eq:convolution}) one can see that
$F(X_{\rm max})$ converges to the exponential form
$\propto e^{-X_{\rm max}/\lambda_p}$ if the factor in square parenthesis
in the right--hand side of the last equation becomes constant for large values
of $X_{\rm max}$. This is true if the function 
$G(Y)$ vanishes sufficiently rapidly for large $Y$,
so that replacing with infinity the upper limit of the integration does not change the result.

This conclusion can be also obtained studying the $X_{\rm max}$ dependent slope $\Lambda (X_{\rm max})$.
Dropping the subscript in the notation for $X_{\rm max}$ one has:
\begin{equation}
 \frac{1}{\Lambda (X)} = - \frac{1}{F(X)} \frac{dF}{dX} = 
 \frac{1}{\lambda_p} - \frac{1}{\lambda_p} ~\frac{G(X)}{F(X)}
= \frac{1}{\lambda_p}
 - \frac{G(X)}{\int_0^X dY ~G(Y)~e^{(Y-X)/\lambda_p}} ~.
 \label{eq:lambda_x}
\end{equation}
For large $X$ the slope $\Lambda_p$ converges to $\lambda_p$ if,
in the limit $X \to \infty$, the last term in the equation vanishes. 
This is the case if the product
$G(Y) \, e^{Y/\lambda_p}$ does not diverge too rapidly for $Y \to \infty$,
a condition that it satisfied in essentially all models for shower development
(see more discussion and one example below).

Equation~(\ref{eq:lambda_x}) has the interesting implication $1/\Lambda(X) < 1/\lambda_p$,
and therefore:
\begin{equation}
 \Lambda (X) > \lambda_p ~.
\label{eq:lambda_limit}
\end{equation}
This inequality follows from the fact that the 
correction term in Eq.~(\ref{eq:lambda_x}) is always negative
because the functions $F$ and $G$ are both probability densities
and can only have positive values, and states
that the depth of maximum distribution of a pure proton composition can never fall 
more steeply than the asymptotic exponential behavior for large $X_{\rm max}$.
A ``flattening'' of the distribution is however possible 
for a composition that includes nuclei, and the value of $X_{\rm max}$ where the
flattening occurs would identify the transition from
a range of $X_{\rm max}$ where nuclei give the largest contribution,
to a range where protons are dominant. 

In this discussion the quantity $\Lambda(X)$ is a differential slope
that changes continuosly with the depth of maximum, reaching asymptotically (from above)
the value $\lambda_p$. 
In practice the experimental studies
for the measurement the $p$--air interaction length 
\cite{Baltrusaitis:1984ka,auger-sigma-2012,Abbasi:2015fdr,Abbasi:2020chd}
have fitted the tail of the depth of maximum distribution 
above a minimum value with simple exponential form.
The slope $\Lambda$ of the fit is then related to the
$p$--air interaction length using an adimensional correction factor $K$:
\begin{equation}
\Lambda = K~\lambda_p ~.
\end{equation}
The correction factor $K$ depends on the range where the fit is performed,
and is always $K > 1$ because of the inequality~(\ref{eq:lambda_limit}),
decreasing toward unity when the fit is performed for larger $X_{\rm max}$ values.
The factor is also model dependent, because the exact form of the
convergence of the slope to $\lambda_p$ is determined by the details of shower development,
encoded in the function $G(Y)$.

The method outlined above to measure the $p$--air cross section
can also be used when the CR flux is formed not only by protons,
but also include nuclei.
This is because protons are the most penetrating of the CR components,
and therefore selecting showers with larger and larger 
$X_{\rm max}$ one also selects a sample of events where
protons give a larger and larger contribution to the distribution. 
In practice of course, this program is possible only if protons 
are a sufficiently large fraction of the CR spectrum.

\subsection{Monte Carlo calculations} 
\begin{figure}[bt]
\begin{center}
\includegraphics[width=12.cm]{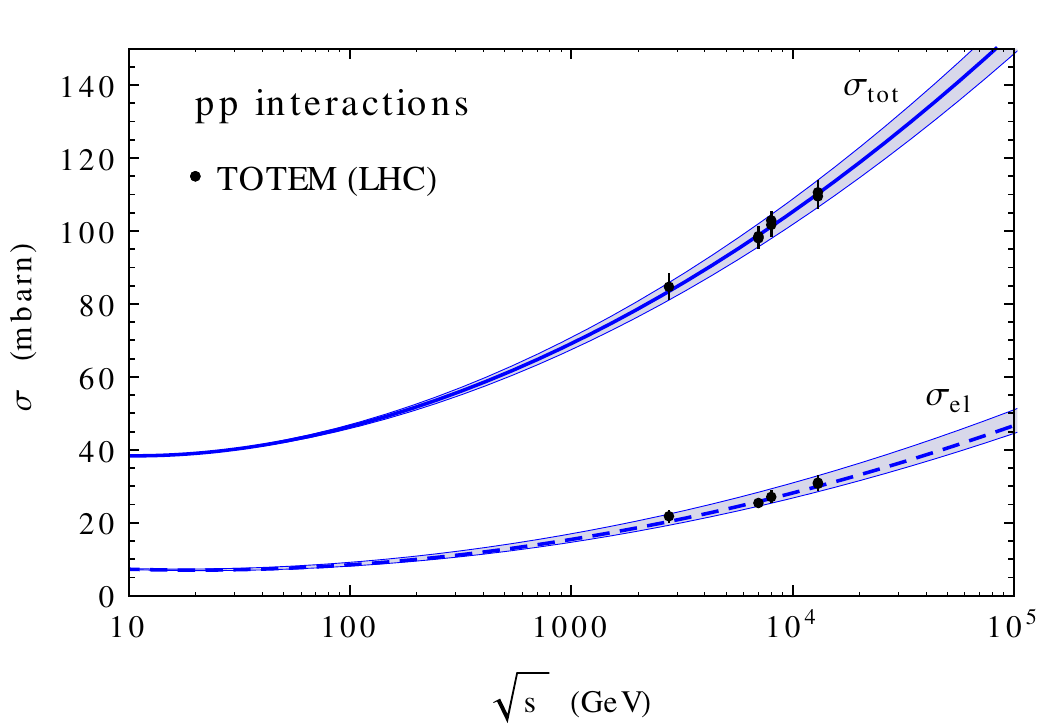}
\end{center}
\caption {\footnotesize
 Total and elastic $pp$ cross sections plotted as a function of the
 c.m. energy $\sqrt{s}$. The points are measurements of the TOTEM
 detector at LHC \cite{Antchev:2013iaa,Antchev:2013paa,Antchev:2017dia}.
 The solid line are fits to the total and
 elastic cross sections that are quadratic in $\log s$ \cite{Antchev:2017dia}.
 The shaded areas are estimates of the uncertainties.
\label{fig:sigma_lhc} }
\end{figure}

To study in more detail the problem of extracting the $p$--air cross section from
cosmic ray observations we have performed
some Monte Carlo simulations, calculating numerically the longitudinal
development of showers generated by very high energy cosmic particles.
For each simulated shower it is then possible to
find the position of the depth of maximum,
obtaining $X_{\rm max}$ distributions with large statistics.
The simulations were performed
for four type of primary particles (protons, {}$^4$He, {}$^{16}$O and {}$^{56}$Fe)
at $E = 10^{18.25}$~eV (approximately the same energy for which
Auger \cite{auger-sigma-2012} has published its measurement of the $p$--air cross section).
The shower development was modeled using the 
Sibyll~2.1 code \cite{sibyll-2.1} to generate the final state of the hadronic interactions,
however the interaction lengths for protons and nuclei adopted to propagate
particles in air were recalculated using Glauber theory \cite{Glauber:1970jm} and
starting from phenomenological fits to the total and elastic $pp$ cross sections
(shown in Fig.~\ref{fig:sigma_lhc}) that are in good
agreeement to the recent measurements at high energy performed at LHC
\cite{Antchev:2013iaa,Antchev:2013paa,Antchev:2017dia}.

\begin{figure}[bt]
\begin{center}
\includegraphics[width=12.cm]{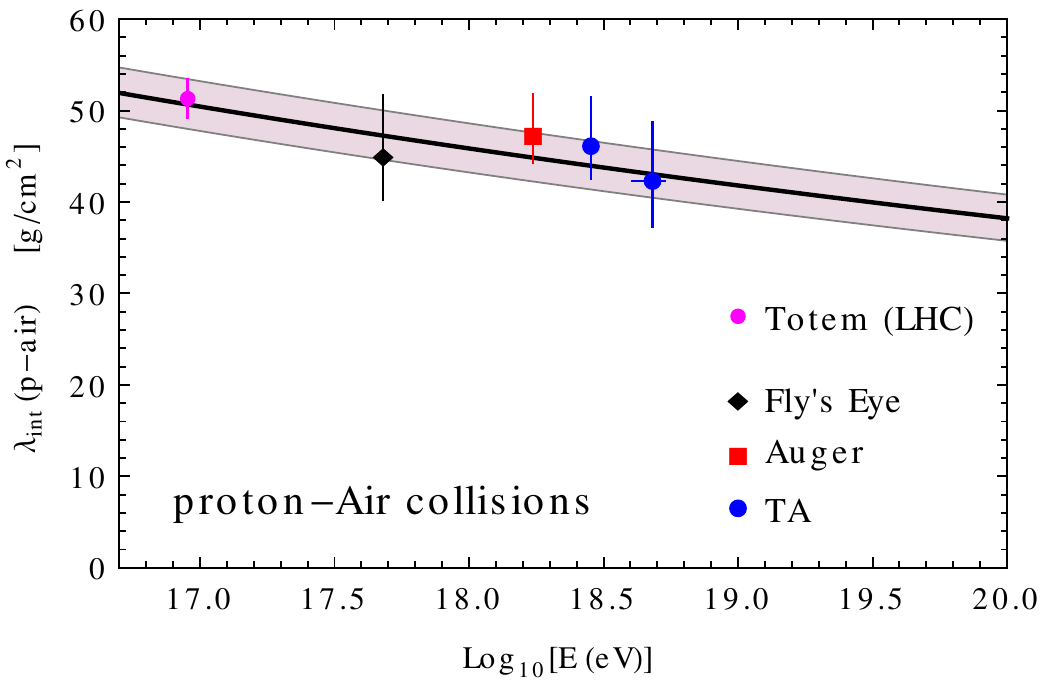}
\end{center}
\caption {\footnotesize
 Proton interaction length in air plotted as a function of the
 projectile laboratory energy. The central line and the shaded area
 are calculated using the best fits
 to the total and elastic $pp$ cross sections and the uncertainties
 shown in Fig.~\protect\ref{fig:sigma_lhc}, and using the algorithms
 of Glauber and Matthiae \cite{Glauber:1970jm} to estimate the proton-nucleus cross sections.
 The points are the estimates of the proton-air interaction length obtained
 from measurements of the longitudinal developments of cosmic ray showers
 by Fly's Eye \cite{Baltrusaitis:1984ka}, Auger \cite{auger-sigma-2012}
 and Telescope Array \cite{Abbasi:2015fdr,Abbasi:2020chd}.
 The lowest energy point is calculated from the measurements at $\sqrt{s} = 13$~TeV
 by TOTEM at LHC \cite{Antchev:2017dia}.
\label{fig:lambda_eas} }
\end{figure}

The resulting $p$--air interaction length is shown
in Fig.~\ref{fig:lambda_eas} together with an uncertainty band
(the shaded area) obtained combining the uncertainties
for $\sigma_{pp}^{\rm tot}$ and $\sigma_{pp}^{\rm el}$ shown in Fig.~\ref{fig:sigma_lhc}.
The uncertainty estimated in this way
is rather small, of order $\approx \pm 3$~g/cm$^{-2}$ for the interaction length, or $\pm 30$~mbarn
($\pm 40$~mbarn) for $E \simeq 10^{18}$~eV
($E \simeq 10^{20}$~eV) for the $p$--air cross section. It must of course be stressed
that this is based on an extrapolation.
In the following we will refer to this model
that combines Sibyll~2.1 with the modified interaction lengths as Sibyll~2.1a.

\begin{figure}[bt]
\begin{center}
\includegraphics[width=16.0cm]{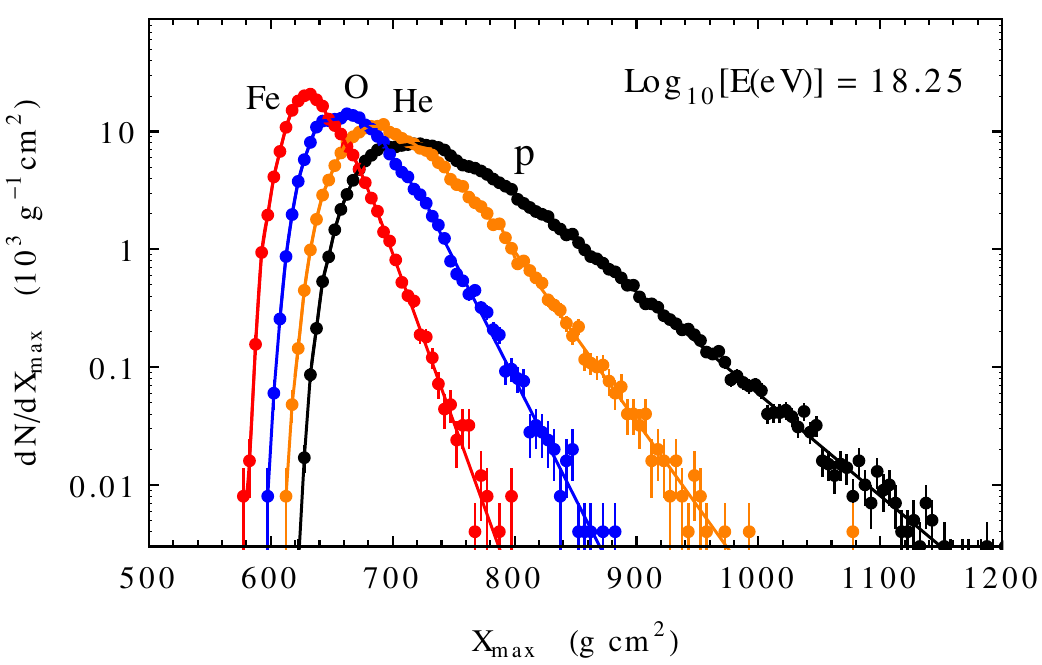}
\end{center}
\caption {\footnotesize
 Distributions of $X_{\rm max}$ for showers generated by particles with energy
 $E=10^{18.25}$~eV.
 The distributions are calculated with Monte Carlo methods
 using the Sibyll~2.1 model and the $p$--air interaction length
 shown in Fig.~\ref{fig:lambda_eas}, 
 for four different nuclei: protons,
 ${}^4$He,
 ${}^{16}$O and 
 ${}^{56}$Fe.
 The high $X_{\rm max}$ part of the distributions has been fitted with a
 simple exponential form: $dN/dX_{\rm max} \propto e^{-X_{\rm max}/\Lambda}$.
\label{fig:nuclei_long} }
\end{figure}

\begin{figure}[bt]
\begin{center}
\includegraphics[height=6.0cm]{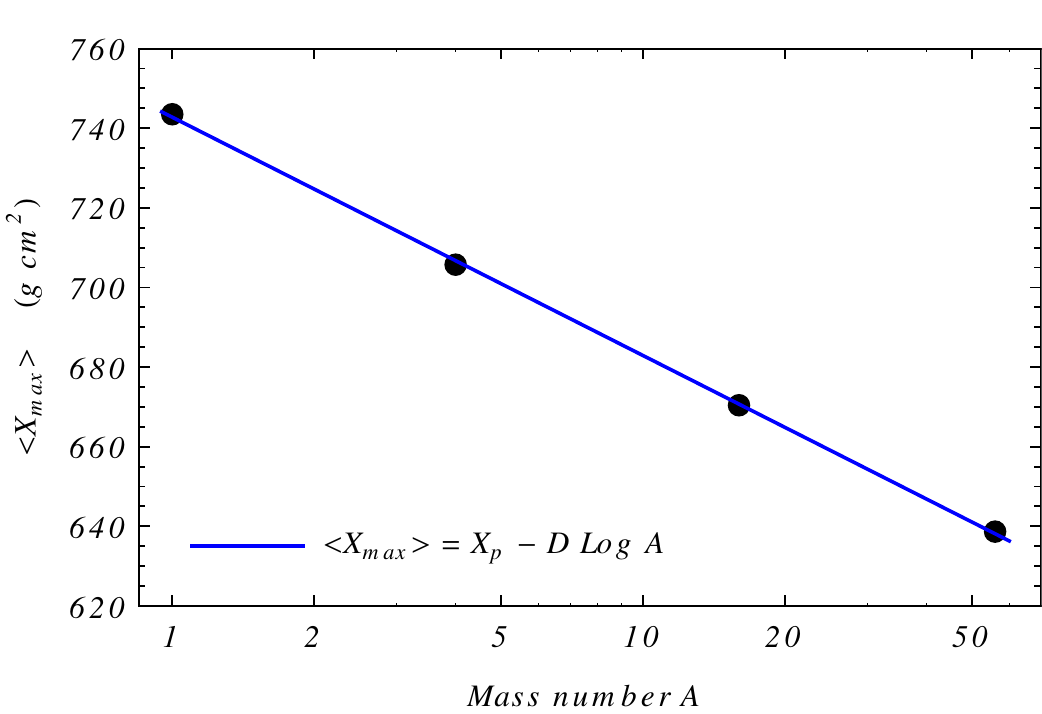}

~~~

\includegraphics[height=6.0cm]{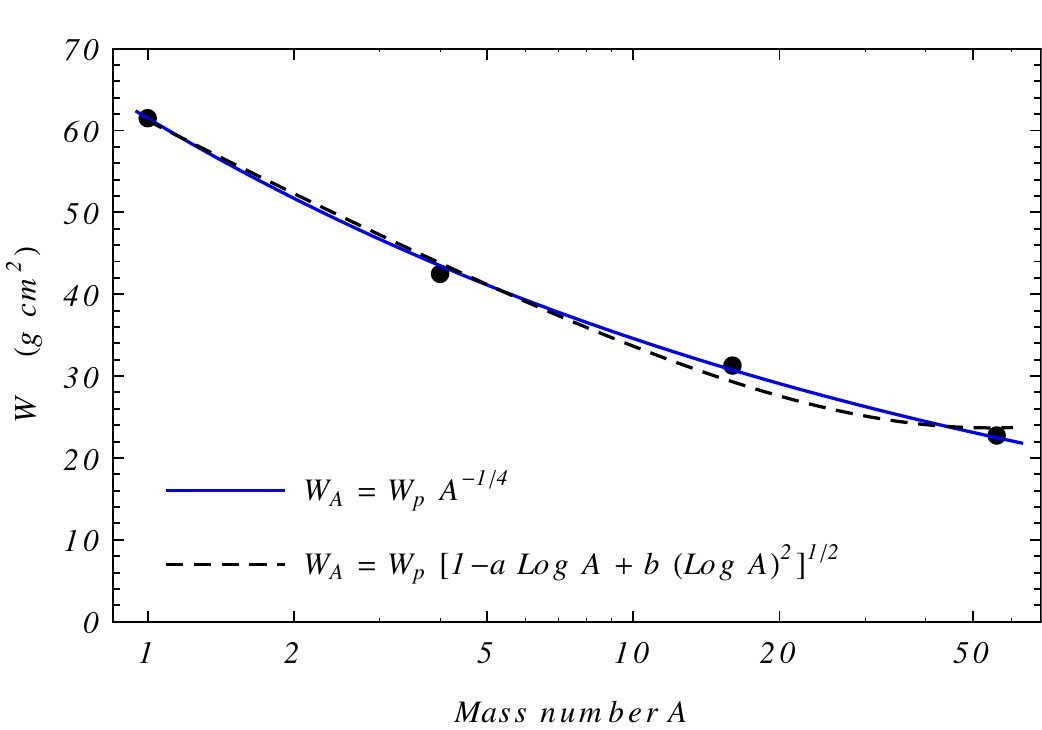}
~~~

\includegraphics[height=6.0cm]{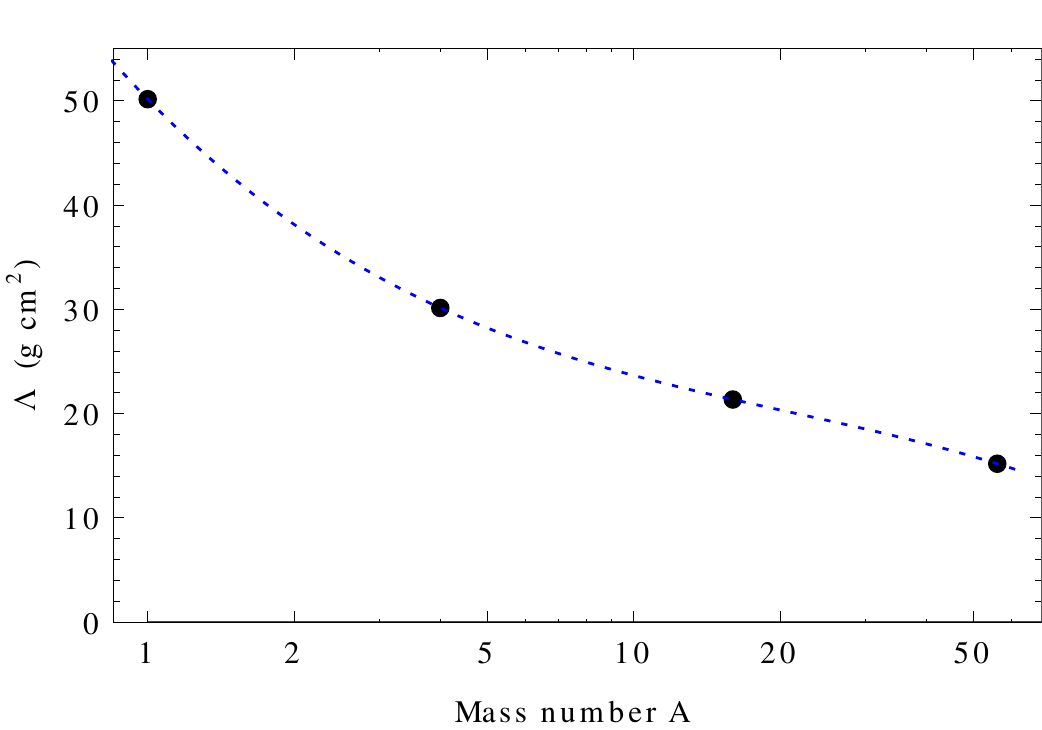}
\end{center}
\caption {\footnotesize
 Parameters of the distributions of $X_{\rm max}$
 for the showers four different nuclei
 (protons, helium, oxygen and iron) at energy $E=10^{18.25}$~eV
 shown in Fig.~\protect\ref{fig:nuclei_long}.
 The top panel shows the average $\langle X_{\rm max} \rangle$, 
 and the line is a linear fit for the relation 
 $\langle X_{\rm max} \rangle = X_0 + D ~\log A$.
 The middle panels shows the r.m.s. of the distributions
 $W = \sqrt{\langle X_{\rm max}^2 \rangle - \langle X_{\rm max} \rangle^2}$,
 and the two lines are analytical approximations of the $A$ dependence
 ($W_A= W_p \, A^{-0.25}$ and $W_A = W_p \, [1- a \, \log A + b \, (\log A)^2]^{1/2}$).
 The bottom panel shows the parameter $\Lambda$ that fits the high energy
 part of the distributions. The line is a polynomial fit to the $\log A$ dependence.
\label{fig:xmax_nuclei} }
\end{figure}

The depth of maximum distributions for the four nuclei
are shown in Fig.~\ref{fig:nuclei_long}.
The averages, widths and also the slopes of exponential fits to the tails
of these distributions are shown in Fig.~\ref{fig:xmax_nuclei}.
Inspecting these results one can observe the following features.
\begin{enumerate}
\item Showers generated by more massive nuclei are less penetrating,
 and the average $\langle X_{\rm max} \rangle$
 (top panel of Fig.~\ref{fig:xmax_nuclei})
 is in good approximation linear in $\log A$
 in agreement with expectations [see Eq.~(\ref{eq:theory_simple})].
 
\item The width of the distributions becomes narrower for larger $A$.
 This is illustrated in the central panel of Fig.~\ref{fig:xmax_nuclei}.

\item For large values of $X_{\rm max}$ the distributions are reasonably
 well described by simple exponentials $F_A (X) \propto e^{-X/\Lambda}$,
 with a slope that depends on the mass number $A$.
 The lines in the figure have (for $A=1$, 4, 16 and 56) 
 slopes $\Lambda = 50.1$, 30.1, 21.3 and 15.2 g/cm$^2$.
\end{enumerate}

As discussed above the shape of the tail of the $X_{\rm max}$ distribution
for protons is related to the proton--air interaction length, that in this calculation
has the value $\lambda_p = 45.9$~g/cm$^2$.
In a Monte Carlo calculation the position of the first interaction point
for each simulated shower is known, and therefore it is possible
to study the distribution of the quantity $Y = X_{\rm max} - X_0$
(that is the depth of maximum measured from the point of first interaction),
and test the validity of Eqs.~(\ref{eq:convolution}) and~(\ref{eq:lambda_x}).

\begin{figure} [ht]
M\begin{center}
\includegraphics[width=12.5cm]{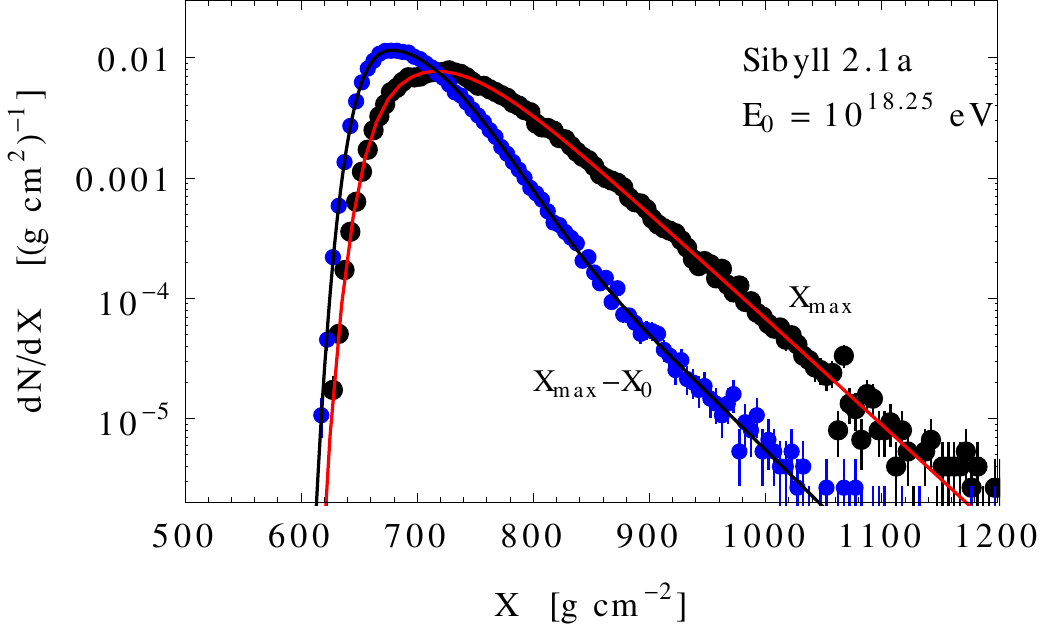}
\end{center}
\caption {\footnotesize
Distribution of $X_{\rm max}$ (bigger points) and
$Y = X_{\rm max}-X_0$ (smaller points) calculated for proton showers
with energy $E_0 = 10^{18.25}$~eV using the Sibyll code and
a shower Monte Carlo model. 
The thin (black) line is a fit to the $Y$ distribution described in the 
main text. The thick (red) line is obtained convoluting the previous 
result with an exponential with slope equal to the
interaction length $\lambda_p(E_0)$.
\label{fig:sib_proton1} }
\end{figure}

\begin{figure} [ht]
\begin{center}
\includegraphics[width=12.5cm]{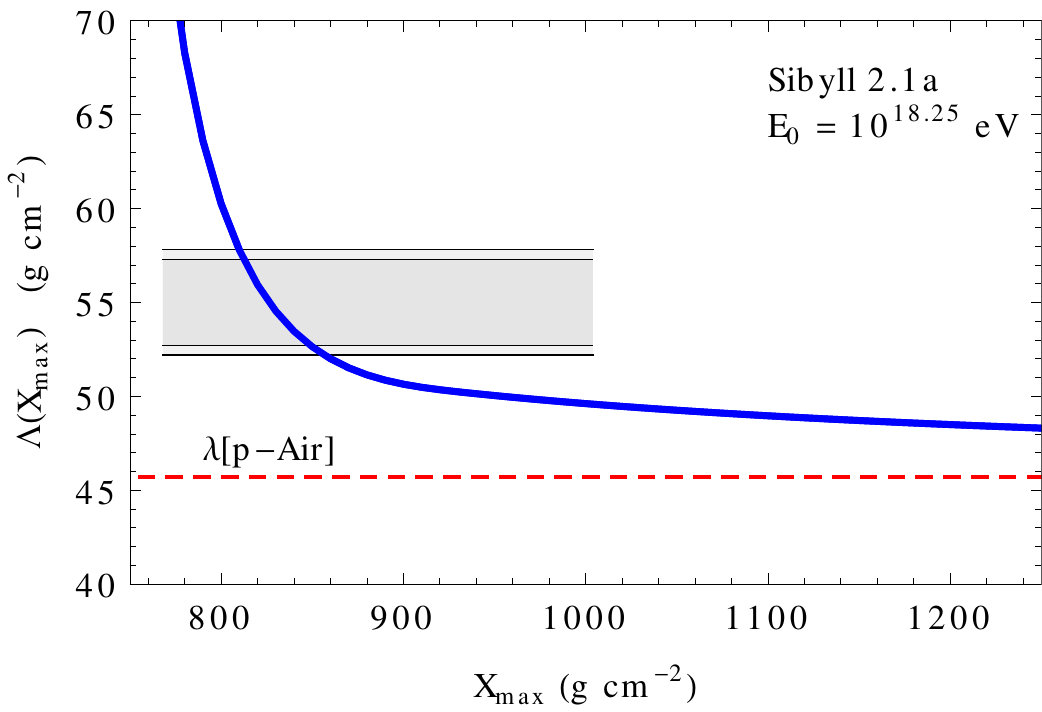}
\end{center}
\caption {\footnotesize
Slope $\Lambda(X_{\rm max})$ of the $X_{\rm max}$ distribution 
obtained with a Monte Carlo calculation for protons
of energy $E_0 = 10^{18.25}$~eV and shown in Fig.~\ref{fig:sib_proton1}.
The dashed line shows the $p$--air interaction length used in the Monte Carlo calculation.
The shaded area shows the range of $X_{\rm max}$ and the best fit value for $\Lambda$
in the study of Auger in \cite{auger-sigma-2012}.
\label{fig:sib_proton2} }
\end{figure}

The distributions of $Y$ and $X_{\rm max}$
for the proton showers are shown in Fig.~\ref{fig:sib_proton1}.
The $Y$ distribution has been fitted
with a smooth curve. The convolution of this curve with
an exponential of slope $\lambda_p$
[see Eq.~(\ref{eq:convolution})] yields a curve
(shown as a red line in the figure) that is a very good description of the $X_{\rm max}$ distribution.
From this expression it is possible to compute
the $X_{\rm max}$ dependent slope $\Lambda(X_{\rm xmax})$,
that is shown in Fig.~\ref{fig:sib_proton2}, where one can see that
the slope approaches from above,
in agreeement with Eq.~(\ref{eq:lambda_limit}),
the value $\lambda_p$ (also shown in the figure as a dashed line).
The convergence of $\Lambda(X_{\rm max}) \to \lambda_p$ is quite slow
due to the fact that the distribution of $G(Y)$
has also an exponential form for large values, with a slope
that is also approximately equaly to $\lambda_p$.
This can be understood noting that the tail of the
$Y$ distribution is formed by events where the final state of
the first interactions contains a ``leading nucleon'' that carries a
large fraction of the initial energy and will then form most of the shower.
Because of this slow convergence, the shape of the tail can in practice be well fitted
with a constant slope in agreement with the ``$K$--factor method'' introduced
by the Fly's Eye collaboration \cite{Baltrusaitis:1984ka}.
This factor is model dependent, because it encodes
the details of shower development, but it is also depends on the $X_{\rm max}$
range where the exponential fit is performed. This range must be chosen
finding a compromise between the need to have a sufficiently high statistics
(a large range), and the desire to have a correction factor closer
to unity (a small range).

\subsection{Comparison with the Auger observations}
\begin{figure}[bt]
\begin{center}
\includegraphics[width=14cm]{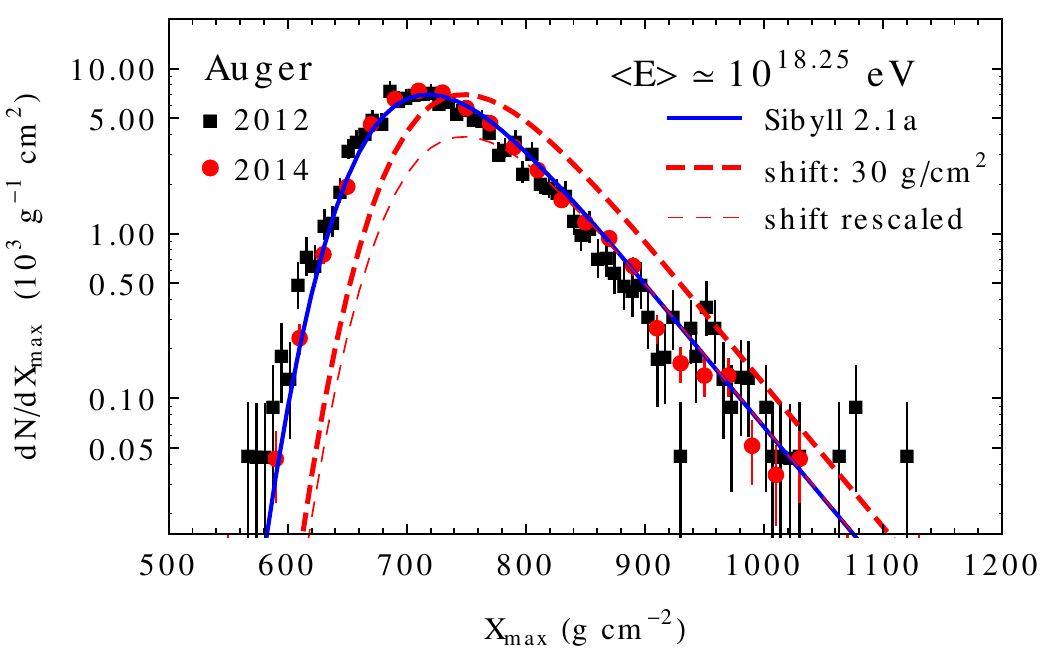}
\end{center}
\caption {\footnotesize
 \label{fig:sigma_18_25}
 The points show the depth of maximum distribution observed by Auger
 for showers with a recontructed average energy around $10^{18.25}$~eV
\cite{auger-sigma-2012} and \cite{Aab:2014kda,auger_data}.
 The thick solid line is the prediction of the Sibyll 2.1a model
 for a pure proton composition. The thick dashed line is the
 same distribution with the showers deeper by 30~g/cm$^2$
 (the difference in average depth of maximum between for showers simulated 
 with the Sibyll 2.1 and Sibyll 2.3c models).
In this case the proton fraction is of order $f_p \simeq 0.55$.
}
\end{figure}

In Fig.~\ref{fig:sigma_18_25} we compare the $X_{\rm max}$ distribution
for protons with the Auger observations after smearing the distribution
with a gaussian resolution with a width of 25~g/cm$^2$.
The Auger data are taken from two sources.
One set of data points is obtained from figure~1 in \cite{auger-sigma-2012},
the work that discusses the measurement of the $p$--air cross section, and refers
to showers observed from December 2004 to September 2010
in the energy range between 10$^{18.0}$ and 10$^{18.5}$~eV,
selected toreduce distortions due to detection acceptance effects.
The second set of points are taken from data tables
publically available online \cite{Aab:2014kda,auger_data}, and
refer to showers observed
from December 2004 to September 2012
in the energy range $10^{18.2}$--$10^{18.3}$~eV.

The comparison of data and simulation is consistent with the conclusions
of the Auger collaboration \cite{auger-sigma-2012}
that obtains for the $p$--air cross section
the value $\sigma_{p{\rm Air}} = 505\pm 22^{+28}_{-36}$~mbarn, that corresponds
to the interaction length $\lambda_p = 47.9\pm 2.1^{+3.7}_{-2.5}$~g/cm$^2$.
This result has been obtained 
fitting the $X_{\rm max}$ distribution in the interval
between 768 and 1004~g/cm$^2$ with an exponential shape, with a best fit slope
$\Lambda = 55.8 \pm 2.3 \pm 1.6$ (where the two errors are statistical and systematic),
and then estimating a correction factor using Monte Carlo calculations.

In our simulation the $p$--air interaction length is set to the value 45.9~g/cm$^2$,
and the shape of the tail of the $X_{`\rm max}$ distribution is consistent with what
is observed by Auger. In Fig.~\ref{fig:sib_proton2} the ($X$ dependent)
slope of the simulation (calculated with good precision using
$10^5$ events) is shown together with the Auger result, showing
reasonably good agreeement.

The main goal of the comparison of the simulation with the data is not to
rediscuss the estimate of the $p$--air cross section obtained by Auger,
but to argue that it is possible to use the study of the
tail of the $X_{\rm max}$ distribution to 
obtain information about the proton fraction in the cosmic ray spectrum.
These results can then be used to constrain the models for shower development.

The measurement of the $p$--air cross section
in fluorescence light detectors is based on the study
the {\em shape} of the {\em tail} of the $X_{\rm max}$ distribution, 
that is fitted with an exponential.
In the cross section study the slope $\Lambda$ of the fit is 
related to the interaction length $\lambda_p$, but the normalization
is discarded (where the normalization is the factor $F_0$ in the exponential fit
$F(X_{\rm max}) = F_0 \; e^{-X_{\rm max}/\Lambda}$, with the distribution $F(X_{\rm max})$
normalized to unity for integration over all values of $X_{\rm max}$).
Also the shape of the $X_{\rm max}$ distribution in the range where
it is not of exponential form is not analyzed.
There are however some obvious merits in
studying not only the shape of the exponential tail of the distribution,
but also its normalization, that accounts for the fraction of events that form it,
and in comparing data and models in the entire $X_{\rm max}$ range.

The interest of such a comparison can be illustrated
inspecting Fig.~\ref{fig:sigma_18_25}. In this figure the data and the Monte Carlo
distributions are both plotted normalized to unity, and one can see that the
exponential tails of the two distributions agree (within errors) both in
shape and in normalization, and in fact that the agreement is reasonably good
for all values of $X_{\rm max}$.
If we make the assumption that the distribution of the data is not distorted by
significant detection acceptance biases, this agreement between
data and simulation indicates that the 
Sibyll~2.1a model can provide a consistent description of the
Auger observations at the energy considered
($E \approx 10^{18.25}$~eV) if the cosmic ray spectrum has a pure proton composition.
The same conclusion can also be reached comparing the values
of $\langle X_{\rm max} \rangle$ and $W$ of the data and of the Monte Carlo
(for a pure proton composition), however the the good matching of the shape
of the distribution add valuable information.

The result on composition is of course model dependent.
A more recent versions of the Sibyll code (Sibyll~2.3c \cite{Fedynitch:2018cbl}, used for comparison
in the Auger analysis discussed above)
predicts that proton showers are on average approximately 30~g/cm$^2$ deeper,
with a distribution of approximately the same shape and width.
For a first order discussion in Fig.~\ref{fig:sigma_18_25}) the Sibyll~2.3c model
is represented shifting by 30 grams the distribution of the older version.
The tail of the $X_{\rm max}$ distribution of a proton spectrum simulated
with the Sibyll~2.3c model has in good approximation the same slope,
but (since the showers are more penetrating), a higher normalization.
It is then possible to match the proton simulation to the Auger data,
but this requires to reduce the proton fraction
by a factor of nearly two (the best fit value is $f_p \approx 0.55$).
The renormalized proton distribution is shown in
Fig.~\ref{fig:sigma_18_25}) as a thin dashed line, and one can immediately see
that this requires the addition of more massive nuclei to the spectrum because
the proton component cannot account for the showers that have small $X_{\rm max}$.

More in general, the slope of the exponential tail of the $X_{\rm max}$ distribution
does offer (if protons are dominant in this range) a measurement of the
$p$--air cross section, that is in good approximation model independent,
but then the matching of the normalizations of the data and of
the proton simulation in the range where the two distributions have an
exponential shape allows to obtain a (model dependent)
estimate of the proton fraction.

Note that the proton fraction cannot be larger than unity, and therefore the 
study of the normalization outlined above
can exclude models where the showers are not sufficiently penetrating.
In fact from the results shown above one can
conclude that models of shower development where
the average depth of maximum is smaller than the Sibyll~2.1a predictions are
strongly disfavored by the observations.

The determination of the proton fraction can
then be combined with the studies of composition based on the
first two momenta of the $X_{\rm max}$ distribution
(and perhaps to other parameters of the distribution, such as the position
of the ``peak'' of the distribution and the position and shape
of its edge for small values $X_{\rm max}$)
to reduce the ambiguities in the estimate of the composition,
and to test the validity of the shower modeling codes.

It might appear that the two programs of 
(i) using the observations of the depth of maximum distribution
for measuring the proton cross section and/or
(ii) use knowledge or theoretical assumptions about
the cross section to measure a (model dependent)
proton fraction, are mutually exclusive.
But this is not the case,
and it is in fact possible to perform these studies simultaneously.
The point is that we have a very robust prediction that the
$X_{\rm max}$ distributions of protons and helium (the lightest nucleus that can
contribute significantly to the CR spectrum) have 
tails of very different shape, with slopes that differ by a factor ($\gtrsim 1.5$)
sufficiently large to allow the identification of a proton component
(or a setting of an upper limit to the proton fraction)
with a reasonably good confidence level.
A slope in the range 45--60~g/cm$^2$ can be safely associated to the
existence of a proton component, and used
(including an appropriate and weakly model dependent correction factor)
to estimate of the $p$--air interaction length.
At the same time the normalization of the distribution in the range where the exponential
form is valid can be interpreted
as a (more strongly model dependent) estimate of the proton fraction.
Such an estimate of the proton--fraction can be them combined with other
observables, such as the average and width 
of the depth of maximum distribution to better constraint the composition.

\subsection{Energy dependence of the proton fraction}
The measurements of the $p$--air cross section obtained from
fluorescence light observations
\cite{Baltrusaitis:1984ka,auger-sigma-2012,Abbasi:2015fdr,Abbasi:2020chd}
span, in terms of laboratory energy, a range (from $10^{17.68}$~eV to $10^{18.68}$~eV)
where there are indications that the CR composition undergoes a significant evolution
(see Fig.~\ref{fig:proton_fraction}), and where the determination
of the composition is strongly model dependent.
A program to study systematically,
as a function of energy, 
the shape of the tail of the $X_{\rm max}$ distribution
with the goal of estimating simultaneously the $p$--air cross section and
the proton fraction could then not only determine more accurately
the $p$--air cross section,
but also give very valuable information on the energy dependence of
CR composition, and so constraining hadronic models.

This study requires to take into account detector acceptance effects
that can generate significant distortions in the experimental distributions,
and cannot be performed here, however for a very preliminary
exploration of the potential of such a program we have
analyzed some publically available data 
of the Pierre Auger Observatory 
for showers with energy larger than $10^{17.8}$~eV
collected from December 2004 to December 2012 \cite{Aab:2014kda,auger_data}.
One example of $X_{\rm max}$ distribution from these data,
for showers in the energy interval
$10^{18.2}$--$10^{18.3}$~eV,
has already been shown in
Fig.~\ref{fig:sigma_18_25},
and compared with the distribution published
in \cite{auger-sigma-2012} where the showers were selected to have
small detection acceptance distortions.
The good agreement between these two data sets
suggest that the detector acceptance effects are not very large.

\begin{figure}[bt]
\begin{center}
 \includegraphics[width=7.0cm]{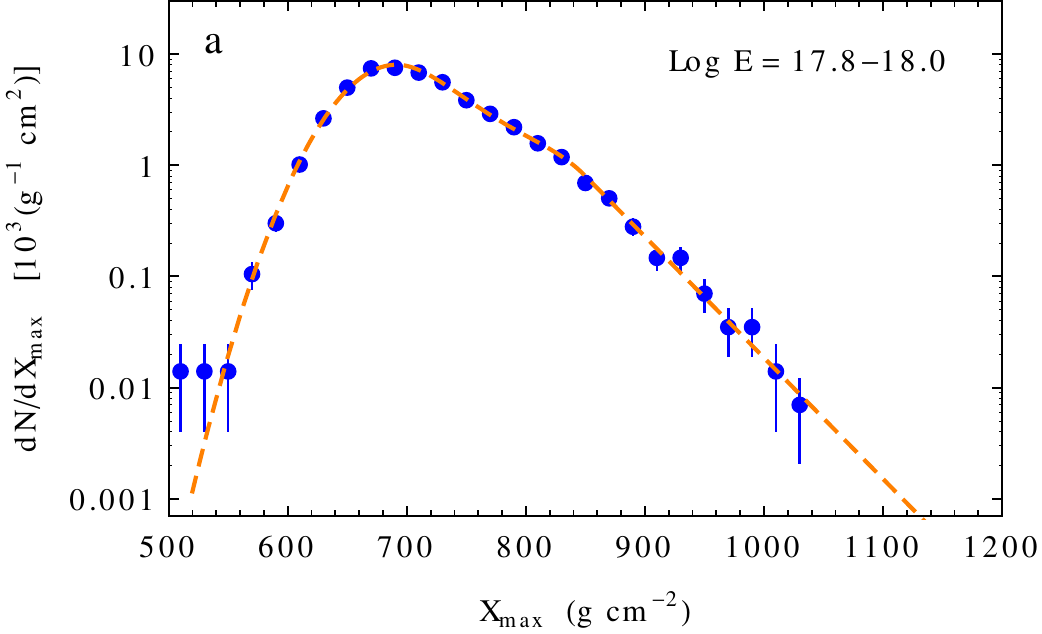}
~~~~\includegraphics[width=7.0cm]{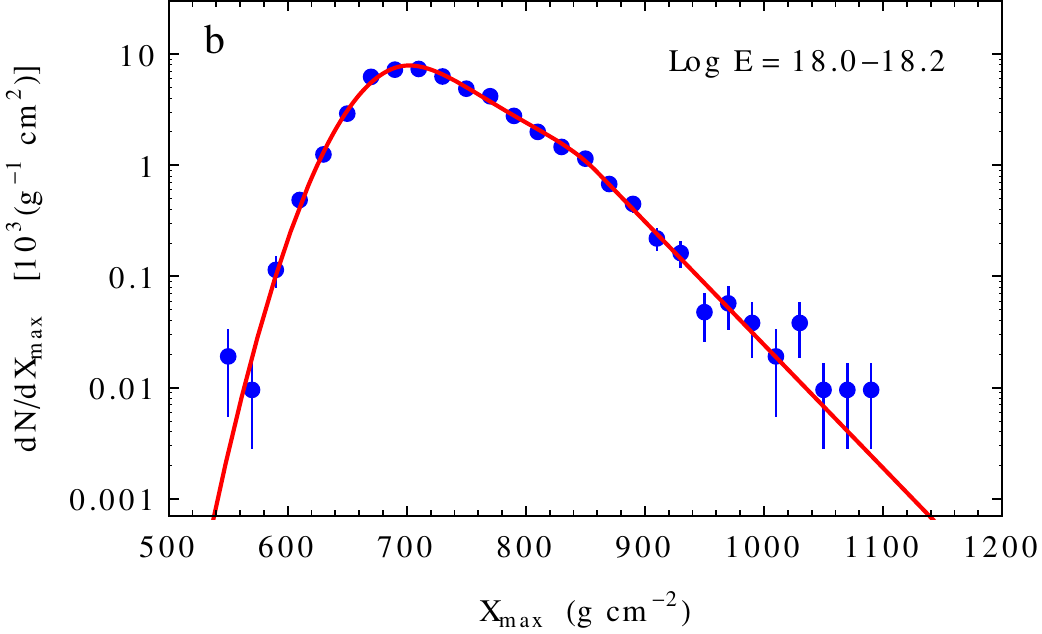}
\end{center}
\begin{center}
 \includegraphics[width=7.0cm]{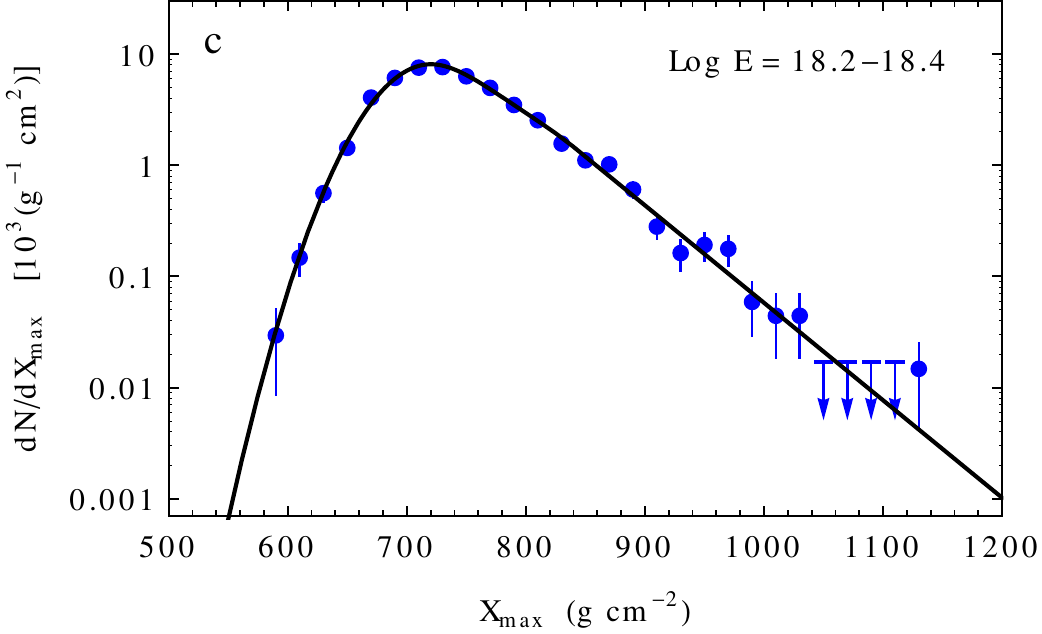}
~~~~\includegraphics[width=7.0cm]{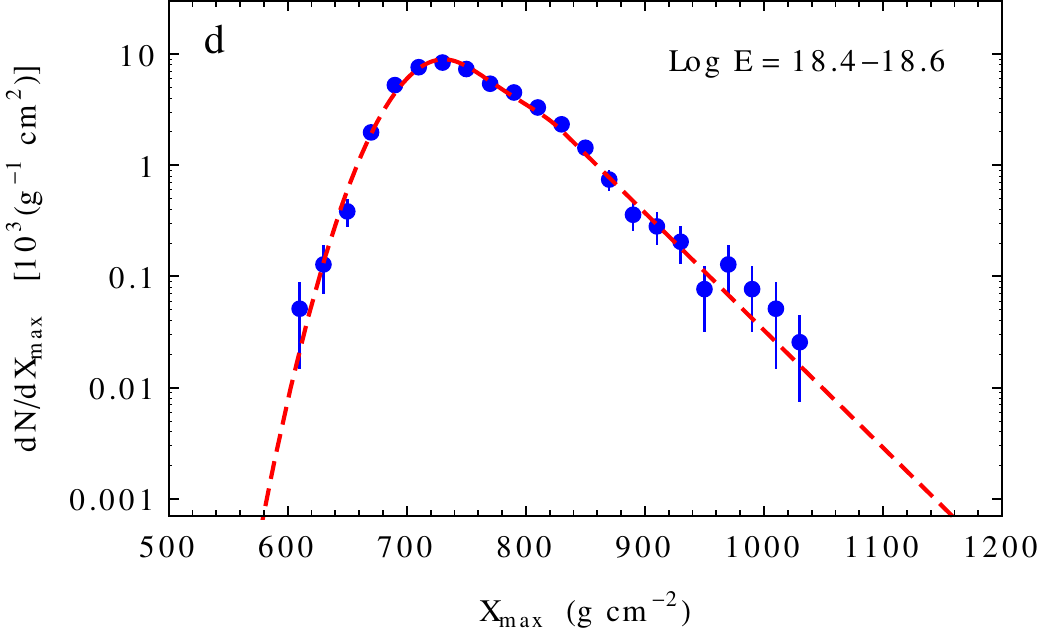}
\end{center}
\begin{center}
 \includegraphics[width=7.0cm]{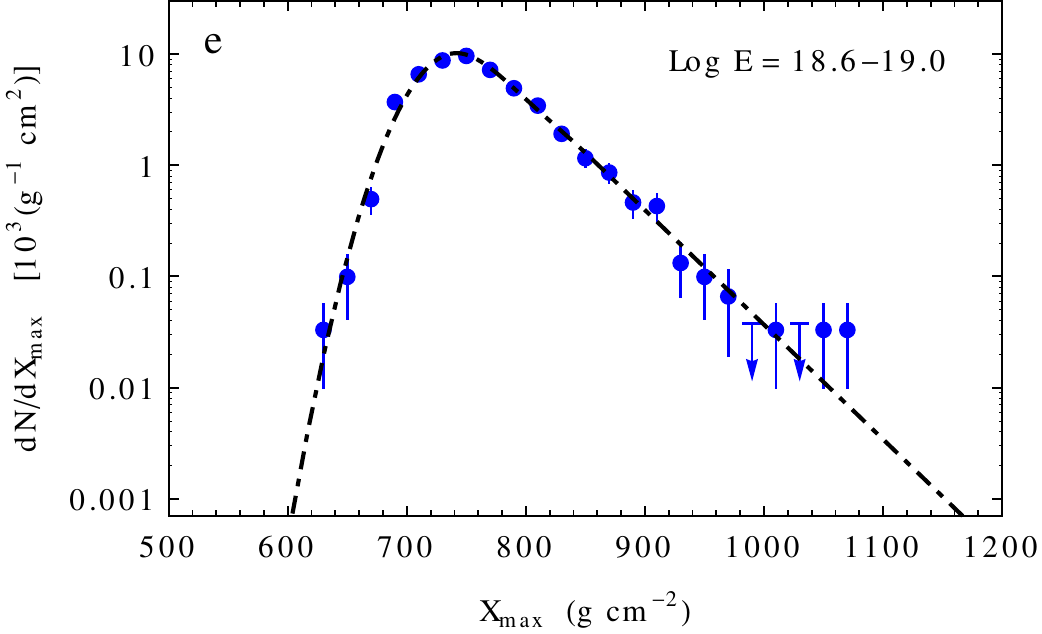}
 ~~~~\includegraphics[width=7.0cm]{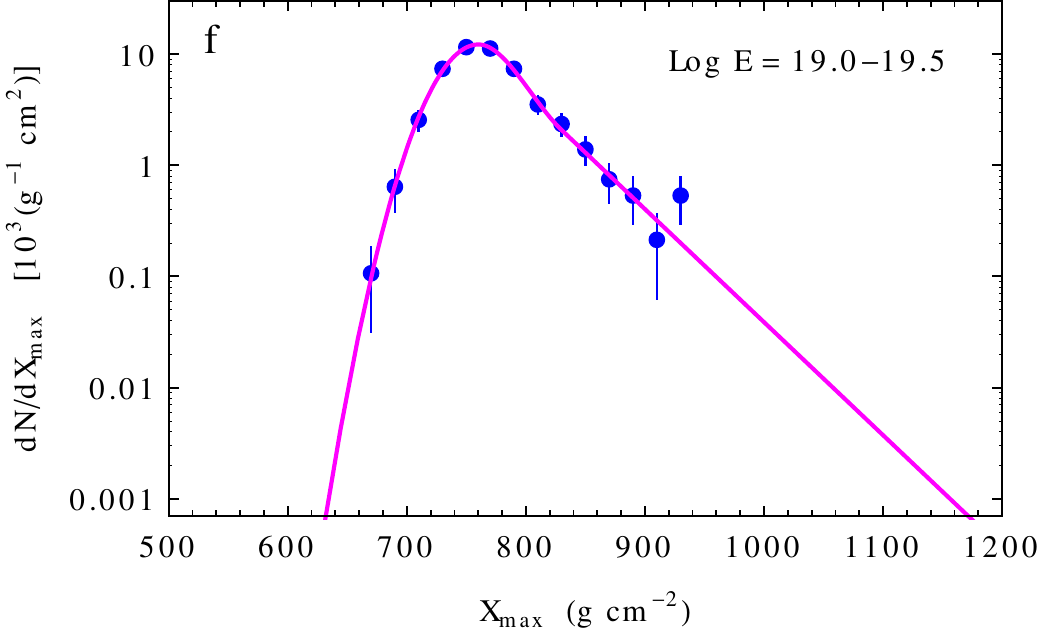}
\end{center}
\caption {\footnotesize
 Depth of maximum distributions of showers detected by Auger
 in different energy intervals \cite{Aab:2014kda,auger_data}.
 The lines are fits discussed in the main text.
\label{fig:fits_all} }
\end{figure}

\begin{figure}[bt]
\begin{center}
 \includegraphics[width=15.0cm]{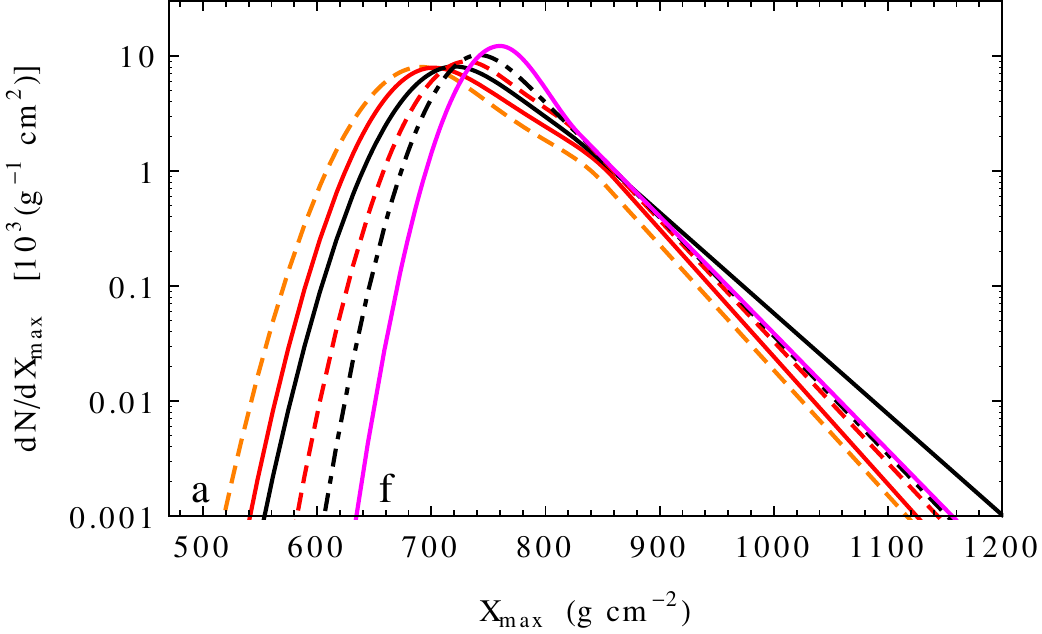}
\end{center}
\caption {\footnotesize
 Fits to the depth of maximum distributions of the CR showers
 detected by Auger \cite{Aab:2014kda,auger_data} in six different
 energy intervals plotted together for comparison.
 Comparisons of the fits with the data are shown in Fig.~\ref{fig:fits_all}, 
 the labeling of the curves (a, $\ldots$, f) is the same as in the panels 
 panels in that figure.
\label{fig:fit_distr} }
\end{figure}

\begin{figure}[bt]
\begin{center}
 \includegraphics[width=8.0cm]{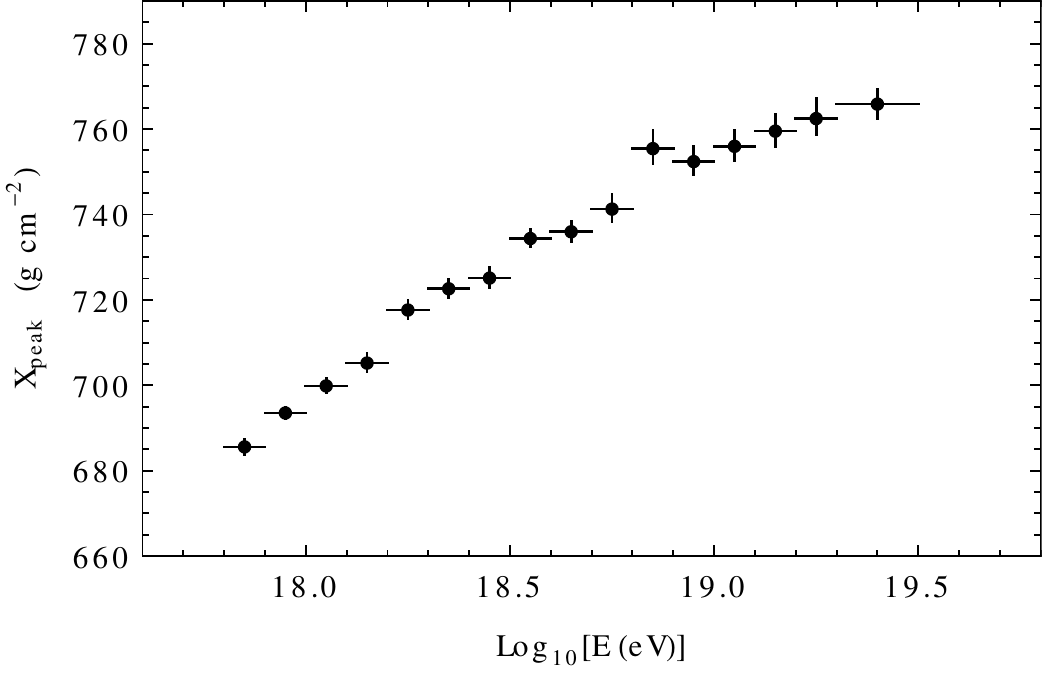}

 ~~~~
 
 \includegraphics[width=8.0cm]{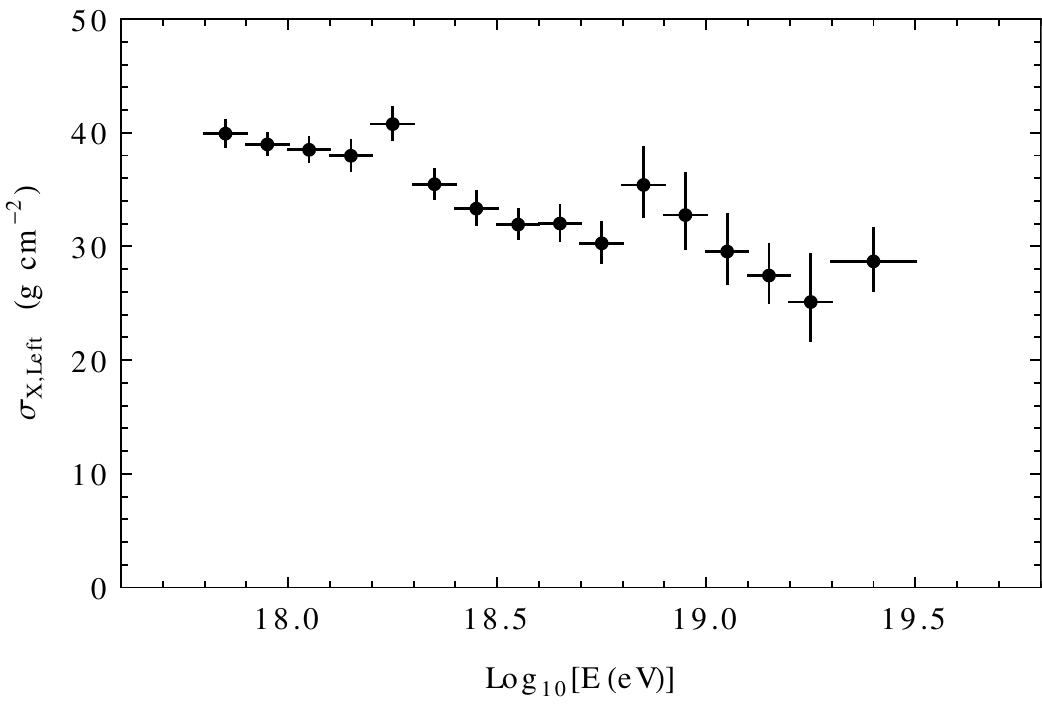}

 ~~~~
 
 \includegraphics[width=7.0cm]{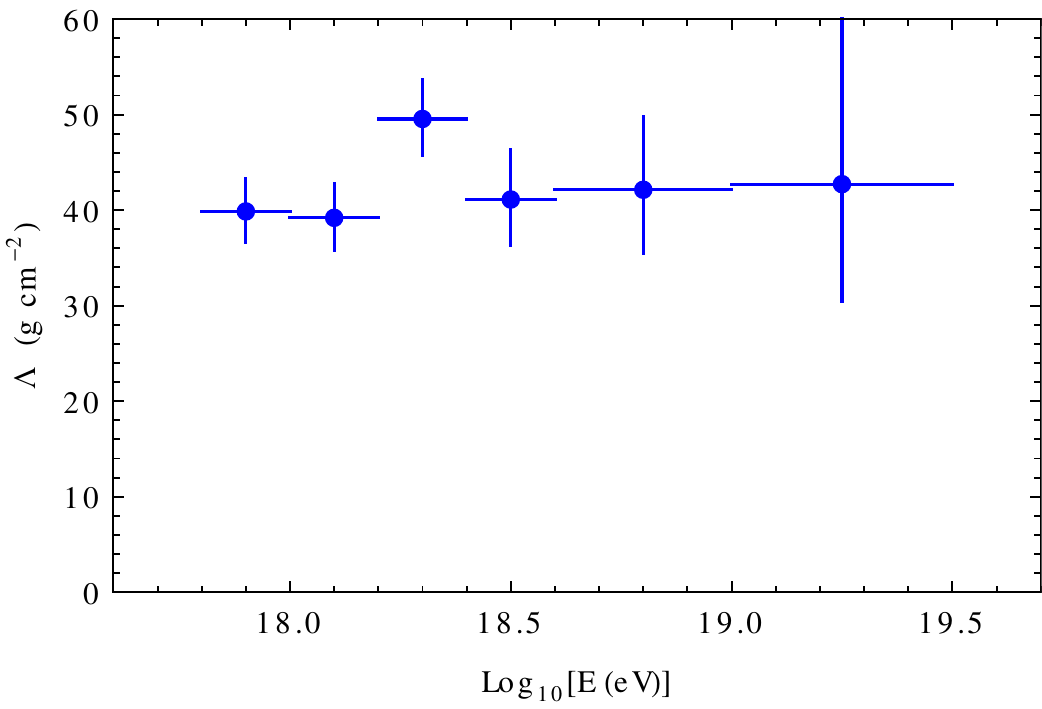}
\end{center}
\caption {\footnotesize
 Best fits parameters for the depth of maximum distributions of the Auger data
 (see Fig.\ref{fig:fits_all}). The top panel shows $X_{\rm peak}$,
 the value of $X$ where the distribution has its maximum value.
 The middle panel shows $\sigma_{X, {\rm left}}$, the width of the Gaussian that fits the
 distribution for $X < X_{\rm peak}$. The bottom panel shows $\Lambda$, the slope
 of the tail of the distribution for large $X$ values. 
\label{fig:fits_parameters} }
\end{figure}

Fig.~\ref{fig:fits_all} show (as histograms) the $X_{\rm max}$ distributions taken
from \cite{auger_data} for six different energy intervals,
together with fits (the lines) constructed 
joining three different functional forms in three $X_{\rm max}$ intervals: \\
(i) For $X_{\rm max} \le X_{\rm peak}$ (with $X_{\rm peak}$ the position of the maximum 
of the distribution) the data is fitted with a gaussian
defined by three parameters:
the position of the maximum $X_{\rm peak}$,
the width $\sigma_{X,{\rm left}}$ and a normalization. \\
(ii) The tail of the distribution 
($X_{\rm max} > X^*$) is fitted with an exponential
$F(X) = K^* \; e^{-X/\Lambda}$.
The parameter $X^*$ is determined from the data as the
broadest range where an an exponential fit is of good quality.
In the six energy intervals considered
the parameters talkes values between 830 and 860~g/cm$^2$.
The quantities $K^*$ and $\Lambda$,
as discussed above, can be
related to the $p$--air interaction length, and to the proton fraction. \\
(iii) The intermediate range ($X_{\rm peak} \le X \le X^*$) is fitted
with the form $F(X) = \exp [P(X)]$ where $P(X)$ a 3rd order polynomial
 in $X$. Three of the four parameters of the polynomial are however 
 determined by the conditions that $F(X)$ is continuous at the two ends of
 the interval, and that the derivative $F^\prime (X_{\rm peak}) = 0$. \\
This form provide a reasonably good quality fit to the data,
and is in fact excellent for the gaussian part
(at small $X$) and for the exponential part (at large $X$).
The six fits are shown together in Fig.~\ref{fig:fit_distr}, and the
energy dependence of three parameters: $X_{\rm peak}$, $\sigma_{X, {\rm left}}$ and
$\Lambda$ are shown in Fig.~\ref{fig:fits_parameters}.

We find that a consistent interpretation of these results is not easy.
In particular the slopes $\Lambda$ obtained in the fits
(shown in the bottom panel of Fig.~\ref{fig:fits_parameters})
cannot be easily interpreted in terms of a $p$--air interaction length.
In the energy interval between 10$^{18.2}$ and 10$^{18.4}$~eV,
that is where Auger has published the cross section measurement 
the slope is consistent with the result published
by Auger. One expects only a weak energy dependence (and a decrease)
that the $p$--air interaction length, however the $\Lambda$ obtained in the fits
at both lower and higher energy are smaller.
This result can perhaps be attributed
to a smaller proton fraction in the flux, but we cannot exclude here
the presence of detector biases.
A more in depth study is required to reach a reasonably form conclusion.
One can however note that the statistical errors
in the data seem sufficiently small to allow an interesting measurement
(and several more years of data taking are available now).

Another very surprising result of our preliminary analysis is 
that the parameter $X_{\rm max}^*$, above which the distribution
is well described by an exponential, is approximately constant
taking values from 830 to 860~g/cm$^2$ in the different
energy intervals considered.
The absolute normalizations of the distributions
at $X_{\rm max} \approx X_{\rm max}^*$ are also quite similar,
an effect that can be clearly seen in Fig.~\ref{fig:fit_distr}
where all the fits (normalized to unity) are plotted together.
Since the slopes of the distributions are are also approximately
constant, this also implies that the fraction of events
in the exponential tail of the distribution changes only little with energy.
For a pure proton composition the models predict
distributions that have exponential tails that account for an approximately
constant fraction of the events, however the position of the tail
covers an $X_{\rm max}$ that grows logarithmically with energy.
In the absence of signficant detection biases these results could be an hint
for the presence of unexpected properties in shower development.

One can also note that is also surprising that the measurement of the proton cross section
obtained by Fly's Eye \cite{Baltrusaitis:1984ka} at $E \simeq 10^{17.7}$~eV
has not been reproduced by new detectors of higher quality,
and that Auger \cite{auger-sigma-2012} and Telescope Array \cite{Abbasi:2015fdr,Abbasi:2020chd}
have measured the proton--air cross sections at different energies.
A comparison of measurements 
obtained at the same energy by more than one detector would
give more confidence in the robustness of the results, and perhaps also be useful
in understanding detector acceptance effects. 

It is interesting to bring attention to the depth of maximum distribution
for the highest energy interval ($10^{19}$--$10^{19.5}$~eV)
shown in the last panel of Fig.~\ref{fig:fits_all}, where there is a
hint of a flattening at $X_{\rm max} \approx 800$~g/cm$^2$.
As discussed above, for a pure proton composition the distribution must always
have a slope larger than the asymptotic value 
[see Eq.~(\ref{eq:lambda_limit})]. 
The flattening observed in the distribution for the highest energy showers
can therefore be considered as a hint for a mixed composition,
with a distribution dominated by a heavier (lighter) component
below (above) the $X_{\rm max}$ of the flattening.

The $X_{\rm max}$ distributions obtained by cosmic ray detectors can be
characterized with several parameters. The Auger studies have focused on
the measurements of the first two momenta ($\langle X_{\rm max} \rangle$ and
$\langle X_{\rm max}^2\rangle$), while the slope $\Lambda$ of the
tail of the distribution has been used to estimate the $p$--air cross sections.
The fits described here define other parameters of the depth of maximum
distribution that can be very useful for a determination of the composition. 
One interesting quantity is $X_{\rm peak}$, the column density
where the distribution has its maximum
(shown in the top panel in Fig.~\ref{fig:fits_parameters}),
and another is $\sigma_{X, {\rm left}}$ that describes the width
of the distribution to the left of the maximum
(middle panel in Fig.~\ref{fig:fits_parameters}).
Together these two quantities determine the position of the small $X_{\rm max}$ edge of the
depth of maximum distribution, that is related (in a model dependent way) to the most
massive component of the CR flux.

The Auger collaboration has already performed studies of the composition
\cite{Aab:2014aea} based on a fit of the entire shape of the depth of maximum
distribution \cite{Aab:2014aea} and not only on the study of the first two momenta.
Such studies determine the fractions of all components in the CR flux,
including the proton one, and are the best and most complete
method to study the composition.
The motivation for a study that focus on the tail of the depth of maximum
distribution is that it is very well suited to test the validity of the models.
Protons are the lighest and most penetrating
component of the CR flux, and therefore, if they give a non negligible 
contribution to the total, they will be emerge as the dominant
contribution of the depth of maximum distribution at high $X_{\rm max}$.
The existence of a proton component can be identified
with a good degree of confidence because the shape of its distribution 
has properties, in particular a large slope of the exponential tail
(related to the $p$--air interaction length),
that are only weakly model dependent.
Other properties, such as the average $\langle X_{\rm max}^{(p)} (E) \rangle$,
or the range where the distribution is exponential, are predicted with
a much larger uncertainty and depend on the description of hadronic
interactions, however information of these properties can be obtained
if and when a proton component is detected.
A systematic study of the proton component
as a function of energy has therefore the potential to determine
(or give stringent constraints) to important quantities such as 
the elongation rate and the average penetration of proton showers.

\section{Conclusions and outlook}
\label{sec:conclusions}
The interpretation of the Auger measurements on the average and width of the
depth of maximum distribution based on current models for shower development,
indicates that the composition of very high energy ($E \gtrsim 10^{17.3}$~eV)
cosmic rays has a surprising energy dependence.
Below the ``ankle'' (at $E \simeq 5 \times 10^{18}$~eV) the composition
is consistent with a mixture of protons and intermediate mass nuclei,
with the proton fraction increasing slowly. These result are consistent
with a CR flux formed by two components of different spectral shape
both of approximate power law form. The origin of these components
does not have a simple explanation.
Above the ankle the composition appears to change very
rapidly with nuclei of larger and larger mass becoming dominant as the energy increases.
This can be interpreted with the hypothesis that the highest energy
sources accelerate particles with a very hard spectrum up to a 
maximum rigidity that in good approximation is equal for all sources.
The mass composition of the particles emitted by these sources is however
quite unexpected.
The implications of these results for high energy astrophysics are profound,
and it is therefore very important to confirm them
with independent measurements, and to validate and strengthen them
with other experimental studies.

It is important to note that the results on composition outlined above
emerge from the comparison of the data with models of shower development that describe 
hadronic interactions using extrapolations of results 
obtained in accelerator experiments at lower energy,
and therefore the possibility that the current models are incorrect,
and that the interpretations based on them
are not valid cannot be at the moment be entirely excluded.
Measurements of the surface arrays (in particular of the muon content
of the showers) have in fact shown
\cite{Aab:2016hkv,Abbasi:2018fkz,Cazon:2020zhx}
that all existing models have flaws and need to be revised.
Experimental studies of the showers that consider different observables
have the potential to shed light on this problem, and
clarify the situation.

Observations of the high energy showers with the fluorescence
technique offer the possibility to identify a proton component in the cosmic ray flux
with a method that can be considered in good approximation as model independent.
The point is that, for a fixed value of the energy,
protons are the most penetrating component of the CR spectrum,
and therefore (if they are present in the flux) will form the tail of 
the depth of maximum distribution for large values of $X_{\rm max}$.
The shape of this tail, in good approximation, takes an exponential form
with a slope that approaches the value of the $p$--air interaction length.
The theoretical prediction based on extrapolations of accelerator
experiments is a slope of order 50~g/cm$^{2}$ for $E \simeq 10^{18}$~eV, that
changes only slowly with energy. The distributions for helium and more massive
nuclei are predicted to have a shape that falls much more rapidly,
(with a slope smaller that $\approx 30$~g/cm$^2$).
This difference allows to identify
the presence of protons in the flux or the setting of an upper limit.
This concept has been used to obtain measurements of the $p$--air interaction length
for laboratory energies between $5 \times 10^{17}$ and
$5 \times 10^{18}$~eV ($\sqrt{s} \approx 30$--95~TeV.
In this work we argue that the identification of a proton component
allows not only to measure the $p$--air cross section,
but also to obtain estimates of the proton fraction and to test the validity of
the shower development models used to interpret the data,
giving information about important properties of the
depth of maximum distribution that depend on the description of hadronic
interactions, such as the elongation rate and the average $\langle X_{\rm max}^{(p)} (E)$.

Measurements of the $p$--air cross sections from fluorescence light observations
have been obtained by different experiments in an energy range that spans one decade
between $5 \times 10^{17}$~eV to $5 \times 10^{18}$~eV.
In this range the Auger observations (interpreted with current models) suggest
that the CR composition is changing, and that the fraction of protons
in the spectrum is not constant.
A systematic study of the proton component in this (or in a broader) energy range
from observations of the tail of the depth of maximum distribution appears therefore
very desirable and has the potential to (i) measure the energy dependence
of the $p$--air cross section, (ii) measure the evolution of the proton fraction,
(iii) estimate the elongation rate (for a pure proton component), (iv) give information
about the average depth of maximum of a proton component and about the width
of the distribution.

The measurements of the $p$--air interaction length obtained by the
fluorescence light detectors have been obtained from a measurement
of the slope of the depth of maximum distribution, without
giving an estimate of the proton fraction. It is however quite obvious
that important information is also contained in the absolute normalization
of the distribution, and in the range of $X_{\rm max}$ where the
the distribution is observed to have exponential form.
These quantities depend on the fraction of protons in the spectrum, and on
the (model dependent) shape of the distributions.

Matching the tail of the depth of maximum distributions of the data and
of a proton Monte Carlo simulation is possible only if the simulation
has the correct $p$--air interaction length.
Many studies and reviews have been dedicated to the modelization of the
cross sections for proton--proton collisions
(see for example \cite{Cudell:2002xe,Pancheri:2016yel,Zyla:2020zbs}).
Combining these studies
with Glauber theory \cite{Glauber:1970jm} (that relates
hadron--nucleon and hadron nucleus interactions), 
the uncertainties on the extrapolation of the $p$--air interaction length
section in the UHECR range are quite small (of order $\sim 3$~g/cm$^2$),
with predictions consistent with the measurements obtained
from CR observations.
Of course, extrapolations can 
be incorrect if new phenomena emerge at higher energy,
and it is therefore of great interest to measure
the $p$--air cross section with UHECR observations
that can reach c.m. energies for nucleon--nucleon collisions
as large as 400~TeV.
Measurements of the $p$--air interaction length
using fluorescence light observations can be performed {\em together}
with a program of estimating the proton fraction in the CR spectrum.
This is possible under the assumption that uncertainties in
hadronic interactions are not so large that it becomes impossible
to identify distributions dominated by protons or helium nuclei.
In this case the $p$--air interaction length can be obtained
(or validated) from the slope of the tail
of the $X_{\rm max}$ distribution in the data,
and then a comparison with a Monte Carlo simulation for proton showers
(that uses the correct cross sections) can be used to estimate the
proton fraction.

Perhaps the strongest motivation to develop studies such as those 
described above, that consider simultaneously several parameters
(such as the average and width, and the slope and normalization of the
large $X_{\rm max}$ tail) of the depth of maximum distributions is that they allow
self--consistency checks that can test the validity of the shower development models,
and also possibly uncover systematic effects in the data taking if they exist.
It should for example be noted that the Telescope Array has obtained one
measurement of the proton cross section at the energy $E \simeq 10^{18.68}$~eV where the
Auger studies suggest that the proton component is already significantly suppressed.
The intriguing tension between these results is perhaps reduced by a difference
between the energy scales of the two experiments in the super ankle range,
but is a question that deserves a careful study.
The very preliminary analysis of the available
Auger data on the depth of maximum distributions
that we have presented in the final part of this work also show some
puzzling features that deserve a more in depth analysis.
In fact, it is puzzling that the Pierre Auger Observatory at the moment has only published
one measurement of the $p$--air cross section at one energy ($E \simeq 10^{18.24}$~eV).
while other detectors with smaller data samples have obtained measurements both
at lower and higher energy, and a systematic study of the proton component as a function
seems very interesting.

More in general and in the same spirit,
it is also very desirable to develop multi--parameter
studies that include even more observables.
For example, from the depth of maximum distribution
one can ontain other interesting quantities such as the most likely
value ($X_{\rm peak}$), of the parameters that describe
the small $X_{\rm max}$ edge of the distribution
(that we have found can be fitted very well by a gaussian) and 
are determined by the components in the spectrum
with largest mass.
In the present work we have also only discussed fluorescence light
observations, and considered only 
one shape parameter ($X_{\rm max}$) for each detected shower, but
more complex analysis are possible.
Very valuable information about the CR composition and about
hadronic interactions are of course encoded in the ground array data.
Extracting this information is a difficult task, but it is of great
importance for making progress in our understanding of the high energy universe.

\vspace{0.5 cm}
\noindent {\bf Acknowledgments}
The first draft of this paper was prepared during
a visit to the Auger site in occasion of the symposium
for the 20th anniversary of the experiment.
I'm very grateful to the organizers for the kind invitation.
I also acknowledge discussions withn
Andrea Addazi,
Xavier Bertou, 
Jose Bellido,
Antonella Castellina,
Ralph Engel,
Lorenzo Perrone,
Viviana Scherini and 
Silvia Vernetto.


\clearpage

\vspace{0.9 cm}

\clearpage


\vspace{0.9 cm}

\vspace{0.9 cm}

\end{document}